\documentclass[]{raa}            

\usepackage{graphicx,times,epsf,subfigure,float,amssymb}             
\usepackage{longtable}

\begin{document}

   \title{Evolutionary Stages and Disk Properties of Young Stellar Objects in the Perseus Cloud
}

   \volnopage{Vol.0 (200x) No.0, 000--000}      
   \setcounter{page}{1}          

   \author{Hong-Xin Zhang\footnote{CAS-CONICYT Fellow}
     \inst{1,2,3,4,10}
   \and Yu Gao
      \inst{5,6}
   \and Min Fang
      \inst{5,7}      
   \and Hai-Bo Yuan
      \inst{3}      
   \and Yinghe Zhao
      \inst{5,6,8}            
   \and Ruixiang Chang
      \inst{9}                  
   \and Xuejian Jiang
      \inst{5,6}      
   \and Xiao-Wei Liu
      \inst{3}      
    \and A-Li Luo
      \inst{1}     
   \and Hongjun Ma
      \inst{5,6}      
   \and Zhengyi Shao
      \inst{9}      
   \and Xiaolong Wang
      \inst{5,6}      
   }


   \institute{National Astronomical Observatories, Chinese Academy of Sciences, Beijing 100012, China \\
       \and 
            Department of Astronomy, Peking University, Beijing 100871, China; {\it hongxin@pku.edu.cn} \
       \and 
            Kevli Institute for Astronomy and Astrophysics, Peking University, Beijing 100871, China \
       \and     
            Chinese Academy of Sciences South America Center for Astronomy, Camino EI Observatorio \#1515, Las Condes, Santiago, Chile
        \and
             Purple Mountain Observatory, Chinese Academy of Sciences, Nanjing 210008, China; {\it yugao@pmo.ac.cn}  \
        \and
             Key Laboratory of Radio Astronomy, Chinese Academy of Sciences, Nanjing 210008, China \
        \and
             Departamento de F\'isica Te\'orica Universidad Aut\'onoma de Madrid, 28049 Cantoblanco, Madrid, Spain; {\it mfang.cn@gmail.com} \             
        \and
             Infrared Processing and Analysis Center, California Institute of Technology, MS 100-22, Pasadena, CA 91125, USA\ 
        \and
             Key Laboratory for Research in Galaxies and Cosmology, Shanghai Astronomical Observatory, Chinese Academy of Sciences, Shanghai 200030, China \     
         \and 
             Current address: Instituto de Astrof\'isica, Facultad de F\'isica, Pontificia Universidad Cat\'olica de Chile, Av. Vicu\~na Mackenna 4860, 7820436 Macul, Santiago, Chile         
   }

   \date{Received~~2015 April 5; accepted~~2009~~month day}

\abstract{
We investigated the evolutionary stages and disk properties of 211 Young stellar 
objects (YSOs) across the Perseus cloud by modeling the broadband optical to 
mid-infrared (IR) spectral energy distribution (SED).\ Our optical $gri$ photometry 
data were obtained from the recently finished Purple Mountain Observatory (PMO) 
Xuyi Schmidt Telescope Photometric Survey of the Galactic Anti-center (XSTPS-GAC).\  
About 81\% of our sample fall into the Stage {\sc II} phase which is characterized by having 
optically thick disks, and 14\% into the Stage {\sc I} phase characterized by 
having significant infalling envelopes, and the remaining 5\% into the Stage {\sc III} 
phase characterized by having optically thin disks.\ The median stellar age and 
mass of the Perseus YSOs are 3.1 Myr and 0.3 $M_{\odot}$ respectively.\
By exploring the relationships among the turnoff wave bands $\lambda_{\rm turnoff}$  
(longward of which significant IR excesses above the stellar photosphere are observed), the 
excess spectral index $\alpha_{\rm excess}$ as determined for $\lambda$  
$>$ $\lambda_{\rm turnoff}$, and the disk inner radius $R_{\rm in}$ (determined from SED modeling) 
for YSOs of different evolutionary stages, we found that the median and standard deviation of 
$\alpha_{\rm excess}$ of the YSOs with optically thick disks tend to increase with $\lambda_{\rm turnoff}$, 
especially at $\lambda_{\rm turnoff} \geq $ 5.8$\mu$m, whereas the median fractional dust 
luminosities $L_{\rm dust}/L_{\star}$ tend to decrease with increasing $\lambda_{\rm turnoff}$.\ 
This points to an inside-out disk clearing of small dust grains.\ Moreover, a positive correlation 
between $\alpha_{\rm excess}$ and $R_{\rm in}$ was found at $\alpha_{\rm excess} \ga$ 0 
and $R_{\rm in} \ga$ 10 $\times$ the dust sublimation radius $R_{\rm sub}$, irrespective of 
$\lambda_{\rm turnoff}$, $L_{\rm dust}/L_{\star}$ and disk flaring.\ This suggests that the 
outer disk flaring either does not evolve synchronously with the inside-out disk 
clearing of small dust grains or has little appreciable influence on the spectral slopes at 
$\lambda$ $\la$ 24$\mu$m.\ About 23\% of our YSO disks are classified as transitional disks, 
which have $\lambda_{\rm turnoff} \geq$ 5.8$\mu$m and $L_{\rm dust}/L_{\star} > 10^{-3}$.\ 
The transitional disks and full disks occupy distinctly different regions on the 
$L_{\rm dust}/L_{\star}$ vs.\ $\alpha_{\rm excess}$ diagram.\
Taking $L_{\rm dust}/L_{\star}$ as an approximate discriminator of disks with ($>$0.1) and 
without ($<$0.1) considerable accretion activity, we found that 65\% and 35\% of the transitional 
disks may be consistent with being dominantly cleared by photoevaporation and dynamical 
interaction with giant planets respectively.\ None of our transitional disks have 
$\alpha_{\rm excess}$ ($<$0.0) and $L_{\rm dust}/L_{\star}$ ($>$0.1) values that would 
otherwise be suggestive of disk clearing dominantly by grain growth.
}

   \authorrunning{H.-X. Zhang, Y. Gao, \& M. Fang et al. }            
   \titlerunning{YSOs in the Perseus Cloud }  

   \maketitle

%
%
\section{Introduction}           
\label{sect:intro}
The formation and early evolution of stars are among the central problems in Astrophysics.\
Young stellar objects (YSOs), which are primarily identified as pre-main-sequence (PMS) stars by the 
presence of infrared (IR) excess arising from circumstellar disks or surrounding envelopes (e.g.\ Allen et al.\ 2004; Greene et al.\ 1994; Lada 1987), have been extensively studied in nearby star-forming regions 
(e.g.\ Taurus: Luhman et al.\ 2010; NGC 1333: Winston et al.\ 2010; IC 348: Muench et al.\ 2007; 
$\sigma$ Ori: Hern\'andez et al.\ 2007; Tr 37: Sicilia-Aguilar et al.\ 2006; NGC2362: Currie et al.\ 2009; 
Lynds 1630N, 1641: Fang et al.\ 2009, 2013).\ Studying the circumstellar environment, either disks or 
envelopes, around YSOs of different mass is essential to understanding the formation of both stars 
and their planetary systems.\ 

YSOs are traditionally categorized into four classes or evolutionary stages based on the 
spectral index $\alpha$ ($d\log(\lambda F(\lambda))$/$d\log(\lambda)$) of their near- to mid-IR spectral 
energy distributions (SEDs; e.g.\  Andr\'e et al.\ 1993; Greene et al.\ 1994; Lada 1987).\ The youngest 
Class 0 objects are only visible in far-IR to submm wavelengths, and they are thought to 
have envelope mass that exceeds the central stellar mass; Class {\sc I} YSOs ($\alpha$ $\geq$ 0.3) 
are characterized by rising mid-IR SEDs, and may be still in an envelope collapse stage but have the central 
stellar mass exceeding the envelope mass; Class {\sc II} YSOs ($-$1.6 $\leq$ $\alpha$ $<$ $-$0.3) 
have SEDs that peak at near-IR wavelengths, decrease at longer wavelengths that is 
much more gradual than what expected for a stellar photosphere, and they agree well with PMS stars 
with circumstellar accretion disks; Class {\sc III} YSOs ($\alpha$ $<$ $-$1.6) have little or no IR excess, 
and are thought to be in the disk dissipation stage with very little or no circumstellar material.\ In addition, 
Greene et al.\ (1994) introduced an additional {\sc Flat}-spectrum class ($-$0.3 $\leq$ $\alpha$ $<$ 0.3) 
which has spectral indices in between Classes {\sc I} and {\sc II}.\

The star formation process is generally accompanied by the formation, evolution and dispersal 
of circumstellar protoplanetary disks which are believed to be the sites of planet formation.\
In particular, optically thick full disks are usually found in the Class {\sc II} YSOs, whereas the 
evolved or anemic optically thin disks are usually identified with the Class {\sc III} YSOs.\
A lot of important information about the evolutionary stages and disk properties of YSOs is encoded 
in the multi-wavelength SEDs (e.g.\ Robitaille et al.\ 2006, hereafter R06).\
For instance, the optical to near-IR bands offer important constraints on the properties of 
the central source, such as the temperature and bolometric luminosity; the near- to mid-IR 
bands provide a crucial constraint on the inner (from a few AU to tens of AU) disk properties; 
the far-IR to submm bands give strong constraints on the mass of disks and envelopes 
(e.g.\ Andrews \& Williams 2005).\
 
As currently the most active site of low- to intermediate-mass star formation within $\sim$300 pc of the Sun, 
the Perseus cloud ($M$ $\simeq$ 4.8$\times$10$^{3}$ $M_{\odot}$; Evans et al.\ 2009) region is an ideal 
laboratory for studying the formation and early evolution of low- to intermediate-mass stars (e.g.\ Bally et al.\ 2008) 
and the circumstellar disks.\ Recently, the $Spitzer$ telescope observations, especially through 
the ``Cores to Disks'' legacy project (c2d; Evans et al.\ 2003), have led to the identification of over 400 
YSOs (mostly Classes {\sc I} and {\sc II}) toward the Perseus cloud.\ In addition, systematic submm  
continuum surveys of the Perseus region with SCUBA (Hatchell et al.\ 2005) and Bolocam/CSO 
(Enoch et al.\ 2006) have led to the confirmation of over 100 protostellar or starless submm cores, 
and about one-third (one-fifth) of these cores were classified as Class 0 (Class {\sc I}) YSOs.\ About 
two-thirds of the Perseus YSOs are associated with the two major young clusters NGC 1333 and IC 348, 
and the remaining YSOs are either associated with other much smaller clouds, such as Barnard 5, 
Barnard 1, L1455 and L1448, or sparsely distributed across the whole Perseus cloud region 
(e.g.\ Evans et al.\ 2009; J$\phi$rgensen et al.\ 2007).\

A systematic investigation of the evolutionary stages and disk properties of the Perseus YSOs 
with the optical-to-IR SEDs is still lacking.\ Moderately deep broadband $gri$ imaging data were recently 
obtained through PMO's Xuyi Schmidt Telescope Photometric Survey of the Galactic Anti-center 
(XSTPS-GAC;  Liu et al.\ 2014; Zhang et al.\ 2013, 2014; Yuan et al. 2015, in preparation).\ 
In this paper, we combined the $gri$ data with the IR data from 2MASS, $Spitzer$, and WISE in 
order to study the physical properties of the central stellar sources, the evolutionary stages and inner 
disk properties of the Perseus YSOs.\ Future spectroscopic data from LAMOST
on most YSOs will further detail and enhance the broad-band characterization offered 
in this paper.\ \S~\ref{sect: samp} introduces the data and YSO sample 
analyzed in this work.\ The color-magnitude diagrams are presented in \S~\ref{sect: cmds}.\  
\S~\ref{sedmod} presents the results from SED modeling, such as the central stellar masses, 
ages, and the evolutionary stages of the YSOs.\ An investigation of the excess dust emission 
and disk geometry parameters, such as the disk inner radii and outer disk flaring, and implications 
on the dominant disk clearing processes are given in \S~\ref{disks}.\ A brief summary of the main 
results in this work is given in \S~\ref{sect: summ}.\


\section{Sample and Data}\label{sect: samp}

\subsection{Parent Sample of Perseus YSOs}
The most recent census of Perseus YSOs was done by Hsieh \& Lai (2013, HL13) using the photometric  
data from $Spitzer$ c2d legacy project (Evans et al.\ 2009), which carried out a wide-field imaging 
survey of five nearby low-mass star-forming clouds (Serpens, Persues, Ophiuchus, Lupus, and Chamaeleon) 
with both IRAC and MIPS instruments onboard $Spitzer$.\ Instead of simply relying on a cut on one or 
two color-color and color-magnitude diagrams to separate YSOs from main-sequence stars and background 
galaxies, HL13 identified YSOs in a multi-dimensional magnitude space.\ In particular, HL13 used data 
from the $Spitzer$ SWIRE survey of the ELAIS N1 extragalactic field (Surace et al.\ 2004) to 
acquire a control sample for background galaxies, and this control sample was used to define the regions 
occupied by galaxies in the multi-dimensional magnitude space.\ The readers are referred to HL13  
for more details about the YSO identification procedure.\

In total, HL13 identified 469 Perseus YSOs over 3.86 deg$^{2}$ covered by the c2d survey.\ 
Adopting a distance of 250 pc for the Perseus cloud, 3.86 deg$^{2}$ corresponds to about 73.6 pc$^{2}$ 
(Evans et al.\ 2009).\ Among the 469 YSOs, 21\% was classified as Class 0/{\sc I} sources, 10\% as 
Class {\sc Flat} sources, 58\% as Class {\sc II} sources, and 10\% as Class {\sc III} sources based on the 
2MASS $K_{s}$ to MIPS 24$\mu$m spectral indices $\alpha$.\ We note that 429 of the 469 YSOs have 
detections in at least 3 IR bands, and thus the identification of these 429 YSOs in the multi-magnitude 
space should be more reliable than the rest 40.\ In the following, the 429 YSOs will be regarded 
as the parent sample, and our subsample selection and analysis will be based on these 429 YSOs.\

\subsection{Data}
\subsubsection{Broadband $gri$ Photometry from XSTPS-GAC}
From the fall of 2009 to the spring of 2011, the XSTPS-GAC observing project carried out an imaging 
survey toward the Galactic anti-center in SDSS $gri$ bands with the PMO's Xuyi 1.04/1.20m Schmidt 
Telescope.\ The survey covers the sky area from RA $\sim$ 45$^{o}$ to 135$^{o}$ and DEC 
$\sim$ $-$10$^{o}$ to 60$^{o}$, plus an extension of $\sim$ 900 deg$^{2}$ toward the M31/M33 
direction.\ With an exposure time of 90 seconds, the survey reaches $r_{\rm lim}$ $\sim$  
19 in the $r$ band at 10$\sigma$ for point sources.\ The astrometry (accurate to $\sim$ 0.1$''$) 
was calibrated against the PPMXL catalog (Roeser et al.\ 2010), and the PSF-fitting photometry 
was calibrated against the SDSS DR8 catalog using the overlapping sky area with an accuracy 
of 2\%.\ Given the importance of optical bands in constraining the properties of central stellar 
sources of YSOs, XSTPS-GAC point sources with S/N $>$ 2 ($r_{\rm lim}$ $\simeq$ 21 mag) 
will be used in this work.

\subsubsection{$Spitzer$ Data from the c2d Project}
As mentioned above, the Perseus cloud has been observed by the c2d project in the $Spitzer$ IRAC 3.6 (IR1), 
4.5 (IR2), 5.8 (IR3), and 7.9 (IR4) $\mu$m and MIPS 24 (M1), 70 (M2), and 160 (M3) $\mu$m bands.\ 
All data, including imagery and point-source photometry (through PSF fitting) catalogs for IRAC, M1 and M2  
were processed and released by the c2d team.\ In this work, we used the HREL (high reliability) source catalog 
provided by the c2d project\footnote{http://irsa.ipac.caltech.edu/data/SPITZER/C2D/}.\

\subsubsection{2MASS and WISE Data}
The $JHK_{s}$ photometry was taken from the 2MASS Point Source Catalog which reaches a 
$K_{s}$-band limiting magnitude of 14.3 mag at 10$\sigma$.\ The Wide-field Survey Explorer (WISE) 
survey mapped the whole sky in four IR broad bands, i.e.\ 3.4 (W1), 4.6(W2), 12(W3) and 
22(W4) $\mu$m, with a 5$\sigma$ limiting magnitude of 16.6, 15.6, 11.3 and 8.0 mag respectively for 
the four bands.\ In this work, we used the ALLWISE Source Catalog\footnote
{http://wise2.ipac.caltech.edu/docs/release/allwise/} which includes enhanced photometric 
sensitivity and accuracy, and improved astrometric precision compared to the 2012 WISE All-Sky Data Release.\

\subsection{Our Working Sample}
In this work, we selected a subsample of 211 Perseus YSOs from the HL13 parent sample.\
The 211 YSOs were selected by cross-matching the HL13 catalog with all the above data sets, with 
a requirement that each source should have $JHK_s$, IRAC or WISE, M1 or W4, and at least 
one optical band available.\ Among the 211 YSOs, 102 have $g$-band detections with S/N $>$ 2, 
151 have $r$-band detections, and 198 have $i$-band detections.\ We point out that 78\% (99\%)
of the $g$-band detections have S/N $>$ 10 (5), 85\% (99\%) of the $r$-band detections have 
S/N $>$ 10 (5), and 94\% (100\%) of the $i$-band detections have S/N $>$ 10 (5).\
In addition, 27 of our sample YSOs have M2 detections.\ Optical photometry of the 211 
YSOs is given in Table \ref{tab1}.\

\begin{figure}
\centering
\includegraphics[width=0.97\textwidth]{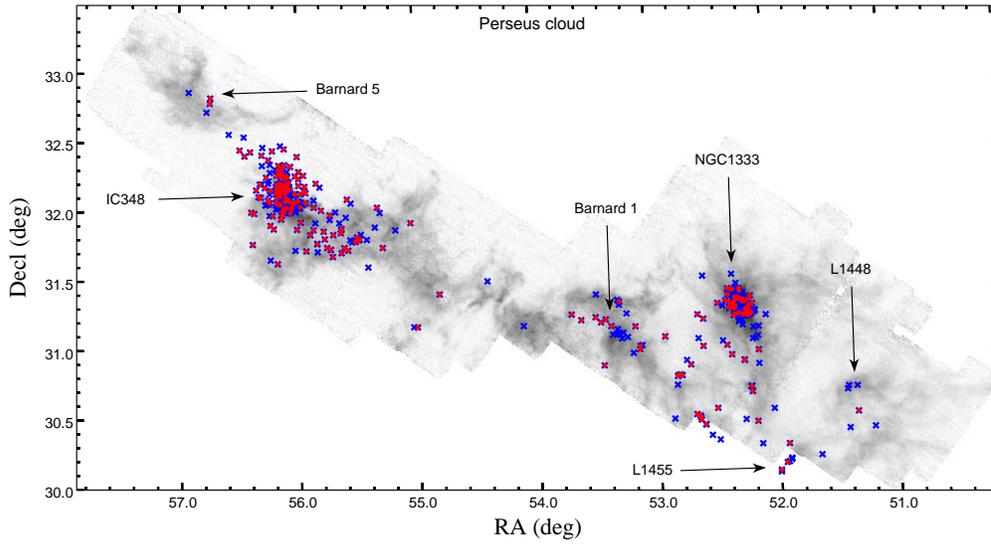}
\vspace{-0.8cm}
\caption{Spatial distribution of the Perseus YSOs is over plotted on the FCRAO 110 GHz $^{13}$CO 
integrated intensity map (greyscale, FWHM $\simeq$ 46'') from the COMPLETE project.\ 
The small red circles mark the 211 YSOs studied in this work, and the blue crosses mark the parent sample 
of 429 YSOs which have at least 3 band detections in the IR wavelength range from 2MASS $J$ to MIPS 24$\mu$m.\
Several well-studied clusters or cores are also annotated in the figure.}
\label{fig1}
\end{figure}

Spatial distribution of the Perseus YSOs is shown in Figure \ref{fig1}, 
the 110 GHz $^{13}$CO integrated intensity map from the The Coordinated Molecular Probe Line 
Extinction Thermal Emission Survey of Star Forming Regions (COMPLETE, Goodman et al.\ 2005; 
Ridge et al.\ 2006) project is overlaid for comparison.\ As already known, most the Perseus YSOs 
are associated with the two major clusters IC 348 and NGC 1333.\ In particular, 83 of the 211 YSOs are 
within 15$'$ ($\sim$ 1.7 pc at a distance of 320 pc, e.g.\ Belikov et al.\ 2002; Evans et al.\ 2009; 
Strom et al.\ 1974; de Zeeuw et al.\ 1999) of IC 348, and 43 are within 15$'$ ($\sim$ 1.3 pc at a distance 
of 250 pc, e.g.\ Evans et al.\ 2009) of NGC 1333.

\begin{figure}
\centering
\includegraphics[width=0.97\textwidth, angle=0]{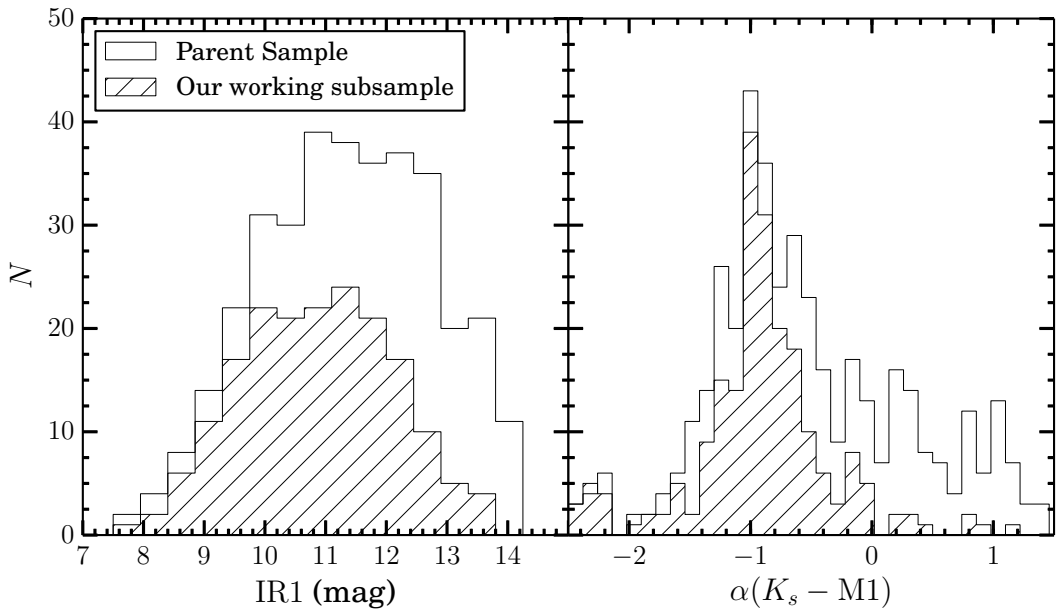}
\vspace{-0.4cm}
\caption{Histograms of the IRAC 3.6$\mu$m (IR1) magnitude ($left$) and spectral index $\alpha$ ($K_s$$-$M1)
($right$) of wavelength ranges from $K_s$ to MIPS 24$\mu$m for the parent sample (open) and our working subsample (hatched) of YSOs.}
\label{fig2}
\end{figure}

Figure \ref{fig2} shows histograms of the IR1 mag and the spectral indices $\alpha$($K_s-$M1) 
for the parent sample and our working subsample.\ 
As can be seen, our 211 YSOs are expected to be statistically unbiased at least at IR1 
$\la$ 10 mag, which would correspond to a stellar mass of $\sim$ 0.9 $M_{\odot}$ at an 
age of $\sim$ 3 Myr for a distance modulus $(m-M)_{0} =$ 7.5 (corresponds to a distance 
of 320 pc for IC 348), according to the PMS evolutionary tracks of Baraffe et al.\ (1998).\ 
Moreover, since hot dust of the circumstellar disks may contribute 
significantly to the IR1 emission, our subsample of YSOs may be statistically unbiased 
down to $M_{\star}$ of slightly below 0.9 $M_{\odot}$.\ In addition, most of our YSOs have 
$\alpha$ $\la$ 0.0, implying that our subsample is dominated by Classes {\sc Flat}, {\sc II} and 
{\sc III} YSOs.\ The spectral index $\alpha$($K_s-$M1) (Evans et al.\ 2009), which 
quantifies the spectral slope from $K_{s}$ to $Spitzer$ 24$\mu$m, was obtained from a 
linear fit to logarithms of all available photometry between $K_s$  and M1.\ Note that for 
sources without M1 data we used the W4 data instead for determining the spectral indices.\

\section{Color-Magnitude Diagrams}\label{sect: cmds}

\begin{figure}
\centering
\includegraphics[width=0.98\textwidth, angle=0]{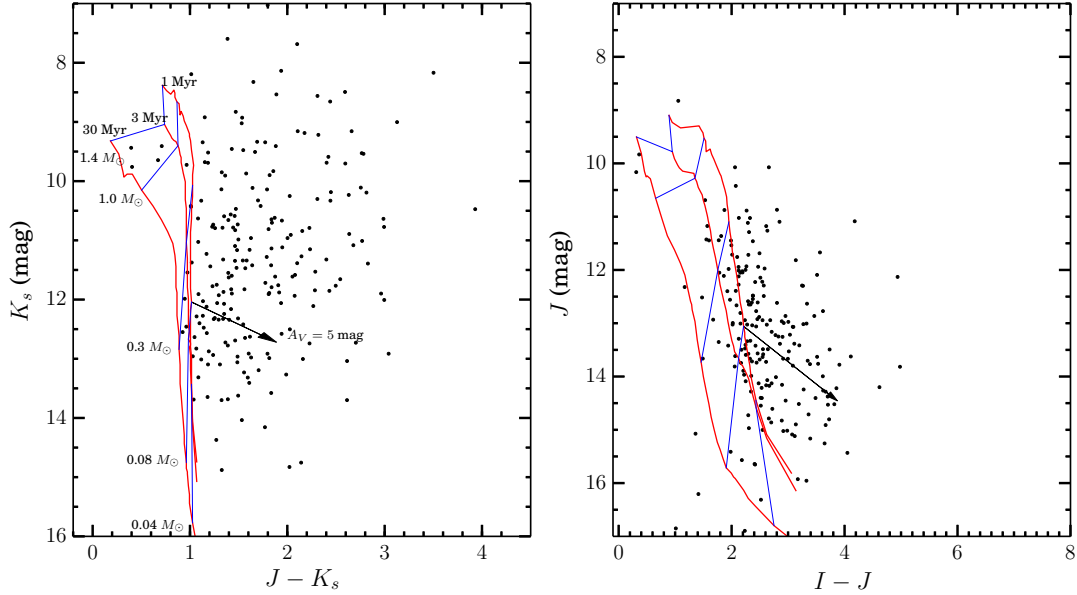}
\vspace{-0.4cm}
\caption{$J-K_s$ vs. $K_s$ ($left$) and $I-J$ vs. $J$ ($right$) color-magnitude diagrams.\
The $left$ panel shows the distribution for all of the 211 YSOs (filled circles) studied in this work, 
and the $right$ panel shows the distribution for 198 YSOs with $i$-band detection.\ 
Overplotted are stellar evolutionary tracks of Baraffe et al.\ (1998) for a stellar mass 
range of $0.02-1.4$ $M_{\odot}$ at three different ages (1, 3, and 30 Myr).\ The black arrow 
in each panel marks the 5-mag visual extinction vector, assuming the Fitzpatrick (1999) extinction 
law with $R_{V}$ = 3.1.\ The evolutionary tracks and extinction vectors shown in the right panel 
are the same as that in the left panel.\ A distance modulus $(m-M)_{0}$ = 7.0 for the Perseus 
YSOs is adopted. 
}
\label{fig3}
\end{figure}

The color-magnitude diagrams for our sample are shown in Figure \ref{fig3}.\
The evolutionary models for low-mass stars and brown dwarfs from Baraffe, Chabrier, \& Allard (1998) 
are also plotted in Figure \ref{fig3} to be compared with our data.\ When plotting the evolutional 
models in Figure \ref{fig3}, the $JHK$ photometry on CIT system as provided by Baraffe et al.\ (1998) 
was transformed to the 2MASS photometric system, and our SDSS $i$ magnitude was transformed 
to the Cousins $I$ magnitude using the transformation equation determined by Lupton (2005).\
The transformation equation of Lupton (2005) involves Cousins $I$, SDSS $r$, and SDSS $i$.\
Among the 198 YSOs that have $i$-band detection, 60 do not have $r$-band detection.\
To put these 60 YSOs on the $I-J$ vs. $J$ diagram (right panel of Figure \ref{fig3}), we adopted 
a median $I$$-$$i$ $=$ $-$0.82, as determined from the YSOs with both $r$ and $i$ detections, to 
transform SDSS $i$ to Cousins $I$.

In Figure \ref{fig3}, most of our YSOs are redder than the pure stellar photosphere 
emission.\ A recent study by Chen et al.\ (2015) found a mean visual interstellar extinction 
of $\la$ 1 mag toward the Perseus region, which is insufficient to explain the red colors of most 
YSOs, especially for their distribution on the $J-K_s$ vs. $K_s$ diagram.\ Therefore, as expected, 
hot dust emission from the inner circumstellar disks of YSOs contributes significantly to the 
$K_s$ band.\ The comparison with theoretical evolutionary tracks implies that the masses of our 
YSOs are mostly above the substellar limit ($\sim$ 0.08 $M_{\odot}$).\ The fact that evolutionary 
tracks at different masses and ages are well separated on the color-magnitude diagram involving 
$I$ band data suggests the importance of optical bands in constraining the properties of the 
central stellar sources of YSOs.

\section{SED Modeling}
\label{sedmod}
\subsection{The Method}
With the broadband SEDs in hand, we used the online SED fitting tool developed by R06 
and Robitaille et al.\ (2007) to extract the relevant physical properties of YSOs 
and their circumstellar disks.\ This online fitting tool offers the possibility of fitting YSO SEDs with a 
precomputed grid of 200,000 synthetic SEDs computed at 10 viewing angles.\ The model SEDs 
account for the contribution from central stellar photosphere emission, circumstellar disks, 
and infalling envelopes.\ In particular, the stellar photosphere emission is modeled with two 
parameters, i.e.\ stellar luminosity and temperature; The disk is treated as a standard flared 
accretion disk and the resultant emission is modeled with six parameters, i.e.\ the disk mass 
($\sim$ 0.001 -- 0.1 $M_{\odot}$), inner radius, outer radius (1 -- 10,000 AU), 
accretion rate, scale height factor and flaring angle; The envelope emission 
is modeled with four parameters, i.e.\ envelope accretion rate, outer radius, cavity density 
(10$^{-22}$ -- 8$\times$10$^{-20}$ g cm$^{-3}$) and cavity opening angle.\ 
In addition, the central stellar masses (0.1 -- 50 $M_{\odot}$) and ages (0.001 -- 10 Myr) were constrained 
by comparing the stellar luminosity and temperature with the PMS evolutionary tracks of 
Bernasconi \& Maeder (1996) and Siess et al.\ (2000).

Before proceeding to the SED modeling for our data, we point out some limitations of the 
R06 SED models (Robitaille 2008) that may be relevant to our current work.\ 
Firstly, the models do not include the case for multiple central stellar sources, which can 
affect the size of the disk/envelope inner holes and thus influence the near- to mid-IR 
emission.\ Secondly, there exist several sets of different PMS evolutionary tracks in the literature, 
besides the Siess et al. tracks as adopted by the R06 SED models, other popular 
PMS tracks include Swenson et al.\ (1994), D'Antona \& Mazzitelli (1997), Baraffe et al.\ (1998), 
Palla \& Stahler (1999),Yi et al.\ (2003), and Dotter et al.\ (2008).\ Adopting different tracks can lead 
to systematic differences on the fitted stellar parameters (e.g.\ Fang et al.\ 2013; Hillenbrand et al.\ 2008), 
and the systematic effects are especially significant for sub-solar mass stars at young ages.\
In particular, uncertainties of age estimation from different tracks for sub-solar mass stars can be 
up to 0.75 dex at young ages ($<$ 10 Myr, Hillenbrand et al.\ 2008).\
Thirdly, the dust opacity law assumed in the models may not be accurate, which would affect the 
determination of disk/envelope accretion and mass.\

When fitting the SEDs, a 5\% absolute flux calibration uncertainty was added in quadrature to the $gri$ 
uncertainties, a 10\% uncertainty was added to $JHK_s$ and IRAC data uncertainties, and a 20\% 
uncertainty was added to the M2 data uncertainties (Evans et al.\ 2009).\ In addition, when both 
IR1, IR2 and W1, W2 data are available, IR1 and IR2 were used in the fitting due to the higher 
resolution of IRAC data.\ An aperture of 10$''$ was used in the fitting.\ In addition, the distance to 
YSOs was allowed to vary from 0.2 to 0.35 kpc, and the foreground interstellar extinction $A_V$ 
was allowed to vary from 0.3 to 30 mag, with the lower limit of $A_V$ being chosen based on the 
Perseus extinction map as determined by Chen et al.\ (2014).\ Besides the 
best-fitting model parameters, all the subsequent well-fit models with reduced 
$\chi^2_{\rm r}$ $-$ $\chi^2_{\rm r,best}$ $<$ 2 were used to define the minimum and 
maximum acceptable physical parameters.

\subsection{The Results}
The range of wavelength coverage determines what physical parameters can be constrained from 
SED modeling.\ A thorough investigation about how the wavelength range of data affects 
the determination of different physical properties of YSOs was given by R06.\
Given our wavelength coverage from optical to MIPS 24$\mu$m (or WISE 22$\mu$m), 
we expect to roughly constrain the central stellar source luminosity, extinction, and the circumstellar disk luminosity.\
Although subject to much larger uncertainties than constraints from spectroscopic data, the central 
stellar masses and ages can still be roughly constrained from broadband SED modeling 
to statistically investigate a large sample, like the one presented in this work.\ 
Moreover, while the masses of the circumstellar disks and envelopes cannot be reliably constrained unless one have 
far-IR to submm data, SED modeling for wavelength ranges short of far-IR can still be used to statistically constrain 
the evolutionary stages of YSOs.\ R06 found that at least 3 different evolutionary stages 
of YSOs can be statistically distinguished based on the fitted stellar masses normalized envelope 
accretion rates ($\dot{M}_{\rm env}/M_{\star}$) and disk masses ($M_{\rm disk}/M_{\star}$).\ 
In particular, the Stage {\sc I} YSOs have significant infalling envelopes and are defined by having 
$\dot{M}_{\rm env}/M_{\star}$ $>$ 10$^{-6}$ yr$^{-1}$; Stage {\sc II} YSOs have 
optically thick disks and are defined by having $\dot{M}_{\rm env}/M_{\star}$ $<$ 10$^{-6}$ yr$^{-1}$ 
and $M_{\rm disk}/M_{\star}$ $>$ 10$^{-6}$; Stage {\sc III} YSOs have optically thin disks and 
are defined by having both $\dot{M}_{\rm env}/M_{\star}$ $<$ 10$^{-6}$ yr$^{-1}$ and 
$M_{\rm disk}/M_{\star}$ $<$ 10$^{-6}$.\ Lastly, the near- to mid-IR SEDs are also 
sensitive to the disk properties, such as the disk inner radius and disk flaring.\ 

\begin{figure}
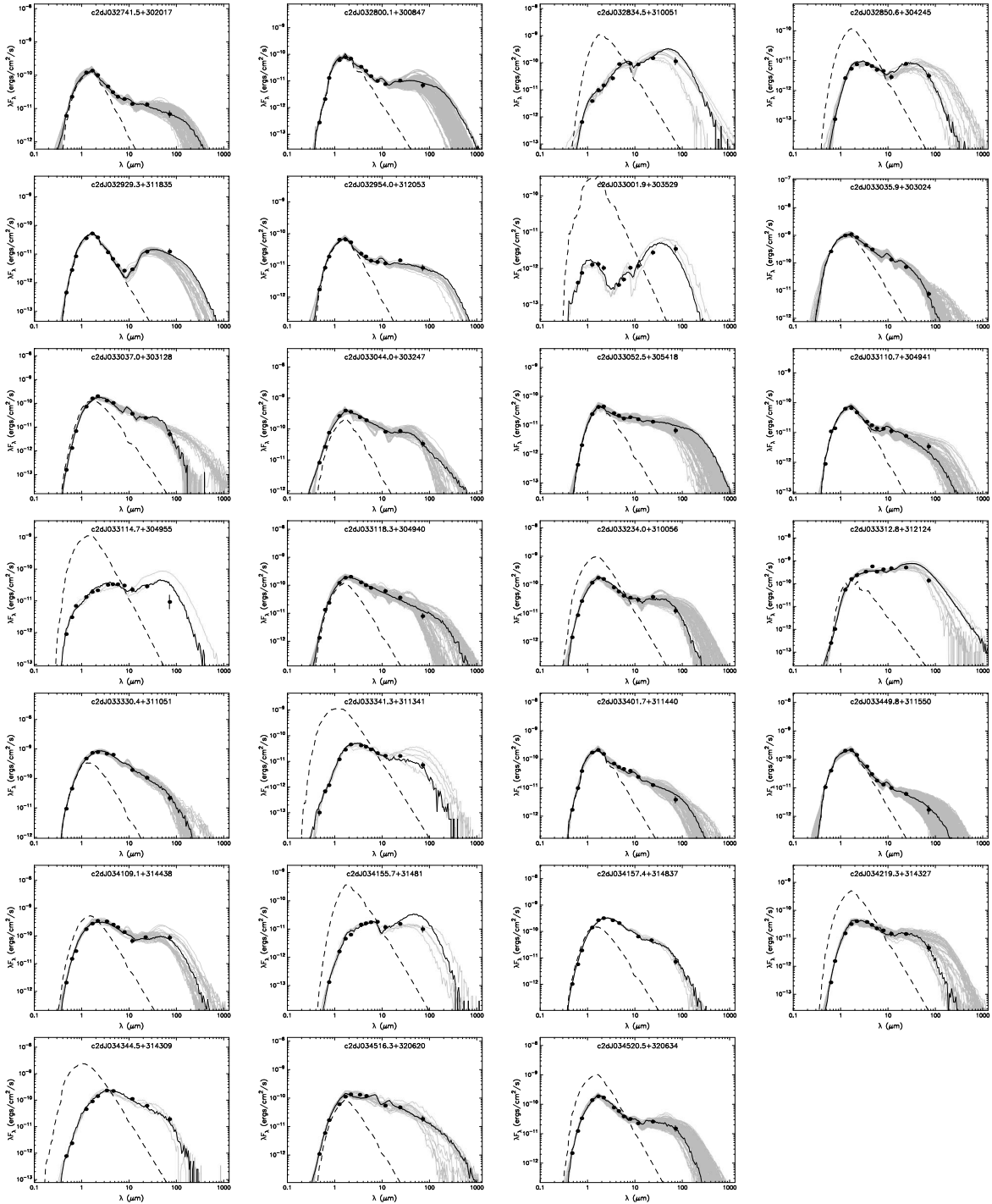

\begin{tabular}[t]{llll}

\includegraphics[width=.245\textwidth,height=0.18\textwidth]{ms2248fig4_1.eps} & 
\includegraphics[width=.245\textwidth,height=0.18\textwidth]{ms2248fig4_2.eps}  &
\includegraphics[width=.245\textwidth,height=0.18\textwidth]{ms2248fig4_3.eps}  &
\includegraphics[width=.245\textwidth,height=0.18\textwidth]{ms2248fig4_4.eps} \\
\includegraphics[width=.245\textwidth,height=0.18\textwidth]{ms2248fig4_5.eps} &
\includegraphics[width=.245\textwidth,height=0.18\textwidth]{ms2248fig4_6.eps} &
\includegraphics[width=.245\textwidth,height=0.18\textwidth]{ms2248fig4_7.eps} &
\includegraphics[width=.245\textwidth,height=0.18\textwidth]{ms2248fig4_8.eps} \\
\includegraphics[width=.245\textwidth,height=0.18\textwidth]{ms2248fig4_9.eps} &
\includegraphics[width=.245\textwidth,height=0.18\textwidth]{ms2248fig4_10.eps} &
\includegraphics[width=.245\textwidth,height=0.18\textwidth]{ms2248fig4_11.eps} & 
\includegraphics[width=.245\textwidth,height=0.18\textwidth]{ms2248fig4_12.eps} \\
\includegraphics[width=.245\textwidth,height=0.18\textwidth]{ms2248fig4_13.eps} &
\includegraphics[width=.245\textwidth,height=0.18\textwidth]{ms2248fig4_14.eps} &
\includegraphics[width=.245\textwidth,height=0.18\textwidth]{ms2248fig4_15.eps} &
\includegraphics[width=.245\textwidth,height=0.18\textwidth]{ms2248fig4_16.eps} \\
\includegraphics[width=.245\textwidth,height=0.18\textwidth]{ms2248fig4_17.eps} &
\includegraphics[width=.245\textwidth,height=0.18\textwidth]{ms2248fig4_18.eps} &
\includegraphics[width=.245\textwidth,height=0.18\textwidth]{ms2248fig4_19.eps} &
\includegraphics[width=.245\textwidth,height=0.18\textwidth]{ms2248fig4_20.eps} \\
\includegraphics[width=.245\textwidth,height=0.18\textwidth]{ms2248fig4_21.eps} &
\includegraphics[width=.245\textwidth,height=0.18\textwidth]{ms2248fig4_22.eps} &
\includegraphics[width=.245\textwidth,height=0.18\textwidth]{ms2248fig4_23.eps} &
\includegraphics[width=.245\textwidth,height=0.18\textwidth]{ms2248fig4_24.eps} \\ 
\includegraphics[width=.245\textwidth,height=0.18\textwidth]{ms2248fig4_25.eps} &
\includegraphics[width=.245\textwidth,height=0.18\textwidth]{ms2248fig4_26.eps} &
\includegraphics[width=.245\textwidth,height=0.18\textwidth]{ms2248fig4_27.eps} \\

\end{tabular}
\vspace{-0.4cm}
\caption{SEDs of 27 Perseus YSOs.\ Among our whole sample, these YSOs have at least one optical band, 
               $JHK_{s}$, IRAC or WISE, MIPS 24$\mu$m or WISE 22$\mu$m, and MIPS 70$\mu$m available 
                    ({\it black points}).\
                    The black solid curve in each panel represents the best-fit model SED of Robitaille et al.\ (2007), 
                    and the grey curves represent all subsequent well-fit models with 
                    $\chi^{2}_{\rm r} - \chi^{2}_{\rm r,best}$ $<$ 2.\
                    The dashed lines illustrate the SEDs of stellar photosphere in the best-fit model, as it would appear 
                    to be in the absence of circumstellar dust.}\label{fig4}
\end{figure}

Figure \ref{fig4} shows the SEDs of the 27 YSOs which have M2-band detections and at the same time 
at least one optical band available.\ The black solid curve in each panel of Figure \ref{fig4} is the 
best-fit model SED, and the grey solid curves represent all subsequent well-fit models with reduced 
$\chi^{2}_{\rm r}$$-$$\chi^{2}_{\rm r, best}$ $<$ 2.\ In addition, SEDs of the best-fit stellar photosphere 
emission (corrected for both the interstellar and circumstellar extinction) are overplotted as dashed curves.\
By calculating likelihood estimator e$^{-\chi^{2}_{\rm r}/2}$ for each well-fit model with $\chi^2_{\rm r}$ 
$-$ $\chi^2_{\rm r,best}$ $<$ 2 for a given YSO, we construct the probability density function (PDF) and 
the correspondent cumulative distribution function (CDF) for parameters such as stellar masses $M_{\star}$, 
ages and disk inner radius $R_{\rm in}$.\ The most probable value for each parameter refers to 
the median of the corresponding PDF, and the confidence interval is defined as covering the central 
95\% of the CDF.\ In what follows in this section, we will present the results for $M_{\star}$,  ages, 
and the evolutionary stages as identified based on the stellar masses normalized disk masses and envelope 
accretion rates.\ Discussion about the disk geometry parameters from SED modeling and fractional dust 
luminosity $L_{\rm dust}$/$L_{\star}$,  where $L_{\rm dust}$ (in units of $L_{\odot}$) is equal to integral 
of the best-fit stellar photosphere subtracted SEDs, will be presented in the next section.\ 
SED modeling results for some relevant parameters, such as $M_{\star}$, ages, and $R_{\rm in}$, 
are listed in Table \ref{tab2}.\

\subsubsection{Stellar Mass Distribution of the Central Stellar Sources}

\begin{figure}
   \centering
   \includegraphics[width=0.97\textwidth, angle=0]{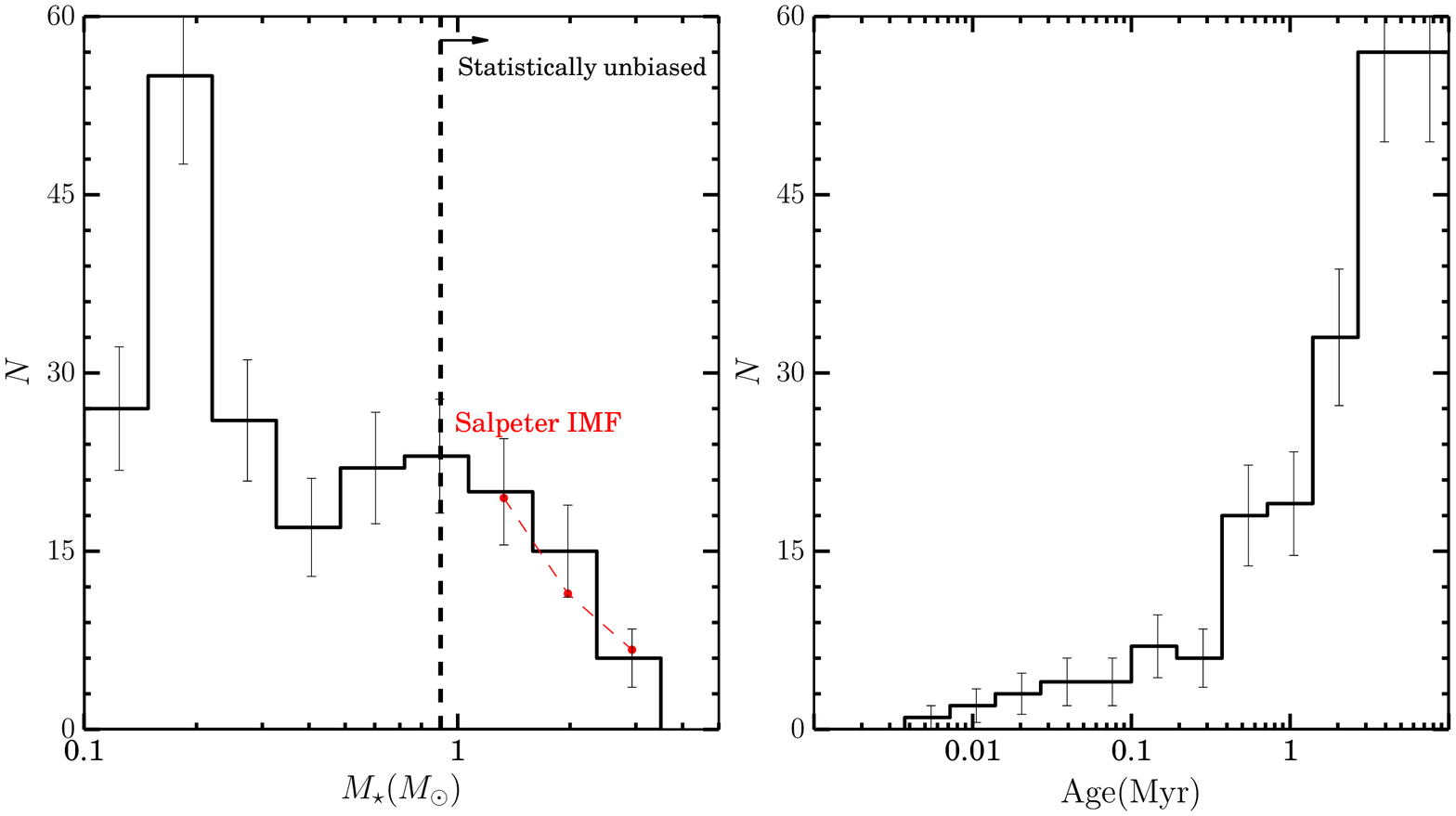}
   \vspace{-0.4cm}
   \caption{Histograms of the stellar masses ($left$) and ages ($right$) for the whole sample of YSOs.\
   The filled red circles in the left panel represents the Salpeter IMF that is scaled  
   to have the same number of observed stars more massive than 0.9 $M_{\odot}$.}
   \label{fig5}
\end{figure}

\begin{figure}
   \centering
   \includegraphics[width=0.97\textwidth, angle=0]{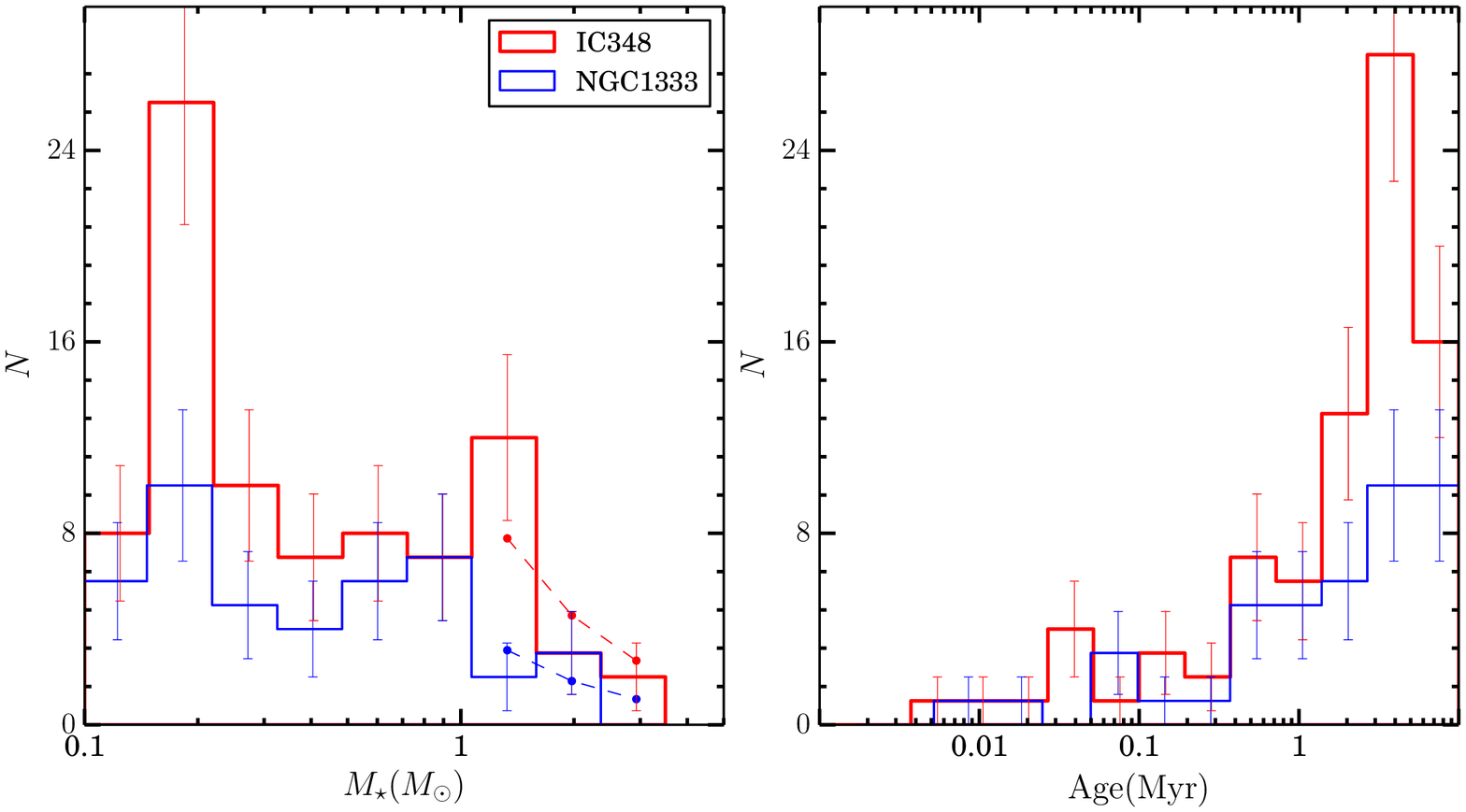}
   \vspace{-0.4cm}
   \caption{Histograms of the stellar masses ($left$) and ages ($right$) for the two major clusters 
   IC 348 (thick red) and NGC 1333 (thin blue).\ YSOs within 15$'$ radius of each of the two clusters 
   are regarded to be associated with the cluster.\ The filled red (blue) circles in the left panel 
   represent the Salpeter IMF for IC 348 (NGC 1333) that is scaled to have the same number 
   of observed stars more massive than 0.9 $M_{\odot}$.}
   \label{fig6}
\end{figure}

Stellar mass histogram of our sample is shown in the left panel of Figure \ref{fig5}.\
As pointed out above, our sample is expected to be statistically unbiased at $M_{\star}$ $\ga$ 0.9 $M_{\odot}$.\ 
We overplot the Salpeter stellar initial mass function (IMF; Salpeter 1955) which was scaled to 
have the same number of stars at $M_{\star}$ $>$ 0.9 $M_{\odot}$ with our YSO sample.\
The error bars in the histogram represent the Poisson noise from number counts.\
It can be seen that the mass distribution of our YSOs at $M_{\star}$ $\ga$ 1 $M_{\odot}$ is 
consistent with the Salpeter IMF within the uncertainties.\ Note that an extended star formation 
history for the Perseus region might make it not straightforward to compare the accumulated 
present-day mass function with the simple Salpeter IMF.\ Although our sample may be 
subjected to significant incompleteness bias below 1 $M_{\sun}$, we note that a flat and broad 
mass distribution from sub-solar to sub-stellar mass limit, as found in our sample, is in 
general agreement with previous studies of low-mass clusters such as IC 348 
(e.g.\ Luhman et al.\ 2003a; Muench et al.\ 2003), NGC 1333 (e.g.\ Wilking et al.\ 2004; 
Greissl et al.\ 2007),Trapezium (e.g.\ Muench et al.\ 2002) and other nearby clusters 
(e.g.\ Andersen et al.\ 2008; Hillenbrand \& Carpenter 2000; Luhman et al.\ 2000; 
Lucas et al.\ 2005; Luhman 2007; Levine et al.\ 2006; Moraux et al.\ 2003; 
Slesnick et al.\ 2004; Scholz et al.\ 2009; Weights et al.\ 2009).\ The median $M_{\star}$ 
of our YSOs is $\simeq$ 0.3 $M_{\odot}$.

Stellar mass distributions of YSOs within 15$'$ radius of each of the two major clusters 
IC 348 and NGC 1333 are shown in the left panel of Figure \ref{fig6}.\ The median stellar masses of 
YSOs in IC 348 and NGC 1333 are $\simeq$ 0.3 $M_{\odot}$.\ 

\subsubsection{Age Distribution of the Central Stellar Sources}
The age histogram for the whole sample is shown in the right panel of Figure \ref{fig5}, and the 
age histograms for each of the two major clusters (again defined with a 15$'$ radius) 
are shown in the right panel of Figure \ref{fig6}.\ The median stellar age of the whole sample is 
$\simeq$ 3.1 Myr, and the median age for YSOs in IC 348 and NGC 1333 is $\simeq$ 2.8 and 2.5 Myr 
respectively.\ A relatively younger age of NGC 1333 than IC 348 is in line with previous studies, and our 
age estimate is consistent with previous studies of YSOs in these two clusters (e.g.\ Herbig 1998; 
Luhman et al.\ 2003b; Lada et al.\ 2006; Winston et al.\ 2009).\

\subsubsection{Uncertainties on Stellar Parameters from SED Modeling}
Determination of the masses and ages of central stellar sources relies on a reasonably 
accurate constraint on the effective temperature $T_{\rm eff}$.\ While it is reasonable to 
statistically explore the distribution of masses and ages determined from broadband 
SED modeling for a large sample, results for individual sources may be subject to large uncertainties.\
In principle, $T_{\rm eff}$ can be accurately constrained by photospheric absorption lines from 
optical or near-IR spectroscopy.\ By comparing our SED-based and the spectroscopy-based 
$T_{\rm eff}$ for 75 IC 348 YSOs that have spectroscopic observations in the literature 
(e.g.\ Luhman et al.\ 2003b; Lada et al.\ 2006; Muzerolle et al.\ 2006; Muench et al.\ 2007), 
we found a median and standard deviation of $T_{\rm eff, SED}$$-$$T_{\rm eff, Spec}$ of $71$ 
($\sim$ 2\%) and 257 K ($\sim$ 7\%) respectively for the 35 objects with $A_{V}$ $<$ 4 mag 
and $T_{\rm eff, Spec}$ $<$ 5000 K, and a median and standard deviation of 
$T_{\rm eff, SED}$$-$$T_{\rm eff, Spec}$ of $68$ ($\sim$ 2\%) and 434 K ($\sim$ 12\%) 
respectively for the 31 objects with $A_{V}$ $>$ 4 mag and $T_{\rm eff, Spec}$ $<$ 5000 K.\ 
In addition, the remaining 9 objects with $T_{\rm eff, Spec}$ $>$ 5000 K have a median and 
standard deviation of $T_{\rm eff, SED}$$-$$T_{\rm eff, Spec}$ of $-1333$ ($\sim$ 24\%) and 
1229 K ($\sim$ 16\%) respectively.\ According to the theoretical evolutionary tracks of Baraffe et al.\ (1998), 
for a PMS star with $T_{\rm eff}$ of 3336 K and an age of 3 Myr, which corresponds to a stellar 
mass of 0.3 $M_{\odot}$, an overestimation of $T_{\rm eff}$ by $\sim$ 350 K ($\sim$ 2--3 subclasses 
in spectral type) at a given luminosity can result in an overestimation of age and mass by factors 
of 3 and 2 respectively.

\subsubsection{Evolutionary Stages of YSOs}

\begin{figure}
   \centering
   \includegraphics[width=0.97\textwidth, height=0.6\textwidth]{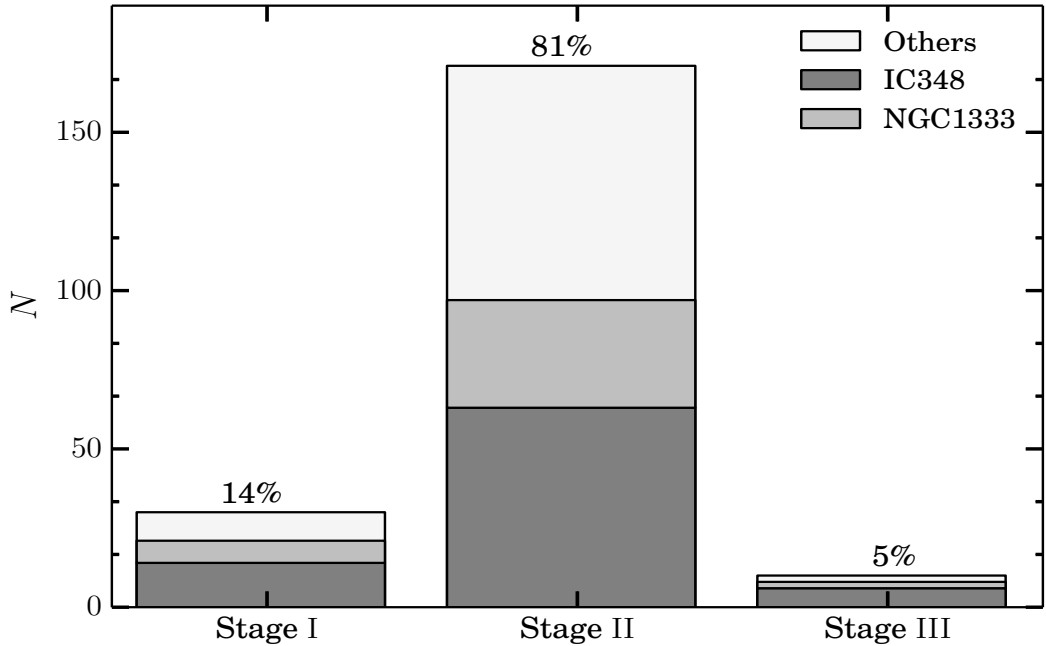}
   \vspace{-0.4cm}
   \caption{Classification of evolutionary stages of our sample based on the fractional envelope 
   accretion rates and disk masses.}
   \label{fig7}
\end{figure}

\begin{figure}
   \centering
   \includegraphics[width=0.97\textwidth, height=0.6\textwidth]{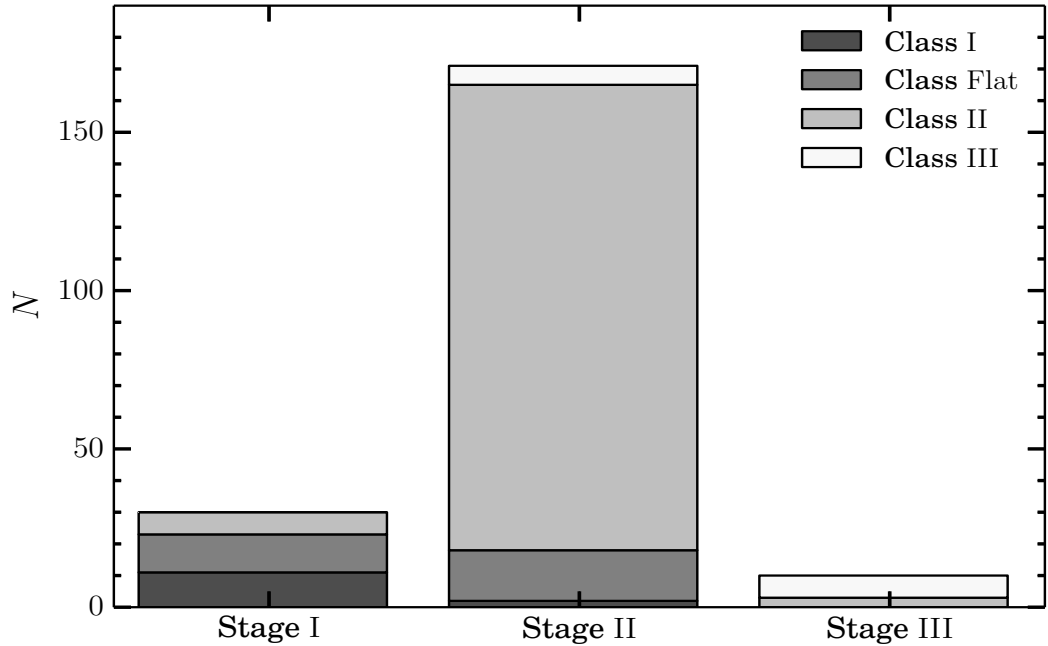}
   \vspace{-0.4cm}
   \caption{Breakdown of different evolutionary stages into different $\alpha$($K_{s}-$M1)--based Classes.}
   \label{fig8}
\end{figure}

Similar to Povich et al.\ (2013), for every YSO, we calculated the accumulated probability 
($P_{\rm stage}$$\propto$$\sum_{\rm model~i}$~e$^{-\chi^{2}_{\rm i, r}/2}$) of it being in 
each of the three Stages (i.e.\ $P_{\rm Stage {\sc I}}$, $P_{\rm Stage {\sc II}}$, 
$P_{\rm Stage {\sc III}}$) based on all the well-fit models with 
$\chi^2_{\rm r}$ $-$ $\chi^{2}_{\rm r,best}$ $<$ 2.\ A YSO is uniquely classified as 
a Stage {\sc I}, {\sc II}, or {\sc III} object if the normalized $P_{\rm Stage}$ $>$ 0.67.\ The result 
of our classification is presented in Figure \ref{fig7}.\ There are 5\% YSOs that can not 
be classified as either Stage {\sc I}, {\sc II}, or {\sc III} if adopting the 0.67 probability threshold.\ 
These 5\% objects were classified as evolutionary stages that have the highest accumulated 
probability.\ The classification of some YSOs into Stage {\sc I} phase may be subject to relatively 
large uncertainties.\ This is because the wavelength coverage of our SEDs is mostly 
limited to $\la$ 24$\mu$m, shortward of which the contribution of excess emission from 
disks dominates over that from the cool infalling envelopes.\ 
Moreover, we note that some of our Stage {\sc I} YSOs with low IR excess luminosities may 
be genuine Stage {\sc II} YSOs with edge-on optically thick disks.\

As can be seen from Figure \ref{fig7}, our sample is dominated by Stage {\sc II} YSOs.\ Moreover, 
the fractions of YSOs in different stages are similar for IC 348, NGC 1333, and the other regions.\
As mentioned in the Introduction section, YSOs have been historically grouped into three or 
four classes based on the spectral index $\alpha$ determined over the wavelength range from $\sim$ 2 
to 20 $\mu$m.\ YSOs of different classes are thought to be in different evolutionary stages (see 
above for references).\ R06 showed that there is a general correspondence between 
the modeling-based ``Stages'' and $\alpha$-based ``Classes'', in the sense that Stage {\sc I} is 
expected to include the Class 0/{\sc I}, Stage {\sc II} is analogous to Class {\sc II}, and Stage {\sc III} to 
Class {\sc III}.\ However, as a set of purely empirical criteria, the Class scheme can be sometimes 
misleading.\ 

Figure \ref{fig8} presents the breakdown of each Stage into different Classes.\
As is shown, a vast majority (94\%) of Class {\sc II} objects are grouped into the Stage {\sc II}, 
and the majority (85\%) of Class {\sc I} objects are grouped into the Stage {\sc I}.\ 
It is noteworthy that the dominant physical Stages for Class {\sc Flat} YSOs are uncertain, 
with about 43\% being in Stage {\sc I} and the remaining 57\% in Stage {\sc II}.\ 
Likewise, the dominant physical Stages for Class {\sc III} YSOs are also uncertain, with 
about 46\% of them being in Stage {\sc II} and the remaining 54\% in Stage {\sc III}.

\section{Properties of the Circumstellar Disks}\label{disks}
The near- to mid-IR excesses above the stellar photosphere level probe the disk properties, 
such as the disk luminosity (basically an integral of the IR excesses) and disk geometry 
(e.g.\ Dullemond et al.\ 2007; Espaillat et al.\ 2013; Hughes et al.\ 2010; Kim et al.\ 2009; 
Mer\'in et al.\ 2010).\ In particular, disk flaring (e.g.\ Kenyon \& Hartmann 1987) and radius 
of the inner disk edge are the two primary disk geometry parameters that shape the SED of IR excesses.\ 
As the disk evolves, dust grains in the inner circumstellar disks may gradually settle down 
(e.g.\ Dullemond \& Dominik 2005) or be cleared out dynamically (Lubow \& D'Angelo 2006) 
or through photoevaporation (Alexander et al.\ 2006a), which leads to a suppression of emission 
excesses progressively from near- to mid-IR wavelengths.\ Features of the IR SEDs that are closely 
related to the disk clearing and flaring include the longest measured wavelength $\lambda_{\rm turnoff}$ 
shortward of which the emission is consistent with being purely from stellar photosphere, and the 
spectral index $\alpha_{\rm excess}$ at $\lambda >$ $\lambda_{\rm turnoff}$ (e.g.\ Cieza et al.\ 2007; 
Harvey et al.\ 2007; Mer\'in et al.\ 2008).\ 

\begin{figure}
   \centering
   \includegraphics[width=0.98\textwidth, height=0.35\textwidth]{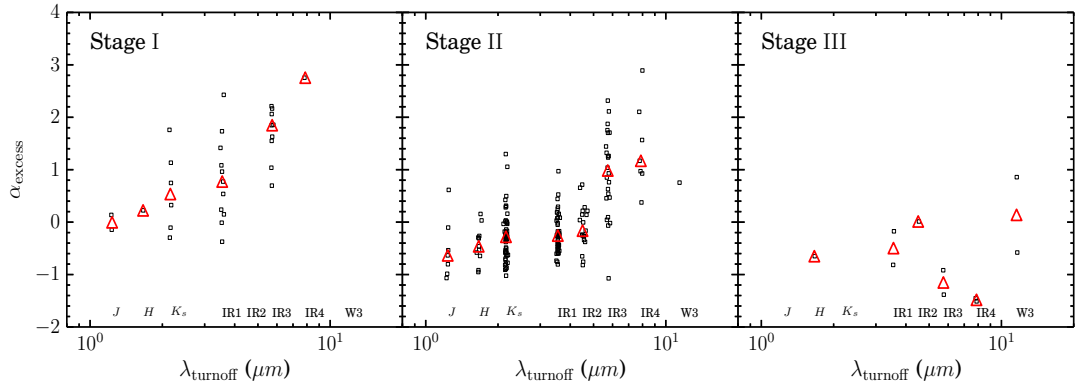}
   \vspace{-0.4cm}
   \caption{Distribution of $\alpha_{\rm excess}$ vs. the wavelength bands $\lambda_{\rm turnoff}$ longward 
   of which IR excesses are observed.\ Stages {\sc\rm I}, {\sc\rm II} and {\sc\rm III} YSOs are shown separately 
   in the {\it left}, {\it middle} and {\it right} panels.\ Median of $\alpha_{\rm excess}$ at each individual 
   $\lambda_{\rm turnoff}$ of different Stages was shown as red triangles.\
   Note that data points at a given wavelength band were slightly shifted randomly in the horizontal direction 
   for clarity.\ See the text for details.}
   \label{fig9}
\end{figure}

\subsection{$\alpha_{\rm excess}$ vs. $\lambda_{\rm turnoff}$}
$\lambda_{\rm turnoff}$ is closely related to the physical scales of the inward disk truncation 
or clearing radius (e.g.\ Calvet et al.\ 2002; Rice et al.\ 2003), and $\alpha_{\rm excess}$ is 
related to both the inward disk clearing and disk flaring which in turn affect the disk temperature 
gradients.\ In particular, for an optically thick disk, a larger spectral index 
corresponds to a shallower temperature gradient (e.g.\ Beckwith et al. 1990).\
By comparing the observed SED of each YSO with the best-fit emergent stellar fluxes (be corrected 
for interstellar extinction), we determined the turnoff wavelength band $\lambda_{\rm turnoff}$,  
longward of which $\geq$ 3$\sigma$ excesses above the stellar photosphere level were observed, and 
calculated $\alpha_{\rm excess}$ for wavelength ranges longward of $\lambda_{\rm turnoff}$.\
Previous studies of YSO IR spectral indices did not exclude the contribution of direct stellar photosphere 
emission.\ In this work, we focus on $\alpha_{\rm excess}$ determined for the photosphere-subtracted 
IR SEDs in order to investigate the disk properties.\ In Table \ref{tab2} we list spectral indices determined 
for both the photosphere-included SEDs ($\alpha_{\rm turnoff}$) and photosphere-subtracted SEDs 
($\alpha_{\rm excess}$) at $\lambda$ $\geq$ $\lambda_{\rm turnoff}$.\ The distribution on 
$\lambda_{\rm turnoff}$ vs. $\alpha_{\rm excess}$ diagram for the subsamples of Stages {\sc\rm I}, 
{\sc\rm II} and {\sc\rm III} YSOs are shown separately in Figure \ref{fig9}.\ 

The median $\alpha_{\rm excess}$ at each $\lambda_{\rm turnoff}$ is also indicated 
as red triangles in Figure \ref{fig9}.\ The majority of Stage {\sc I} YSOs have $\alpha_{\rm excess}$ 
$\ga$ 0.0, whereas the majority of Stage {\sc III} YSOs have $\alpha_{\rm excess}$ $\la$ 0.0.\
Compared to the Stages {\sc I} and {\sc III} YSOs, the Stage {\sc II} YSOs have a larger range of 
$\alpha_{\rm excess}$ from $\sim$ $-1$ to 3.\ The median $\alpha_{\rm excess}$ 
gradually increases with increasing $\lambda_{\rm turnoff}$ for both Stages {\sc I} and {\sc II} 
YSOs.\ No obvious trend of the median $\alpha_{\rm excess}$ with $\lambda_{\rm turnoff}$ 
is found for the Stage {\sc III} YSOs.\ In addition, there is a hint that the standard deviation of 
$\alpha_{\rm excess}$ increases with increasing $\lambda_{\rm turnoff}$ for the Stage {\sc II} 
YSOs which have the largest sample size.\ In particular, the standard deviations of 
$\alpha_{\rm excess}$ for the Stage {\sc II} YSOs of different turnoff wavelengths at $\lambda_{\rm turnoff}$ 
$\leq$ IR2 and $\geq$ IR3 are $\sim$ 0.4 and 0.8 respectively.\
A smaller spread of $\alpha_{\rm turnoff}$ at shorter $\lambda_{\rm turnoff}$ has been 
observed before (e.g.\ Cieza et al.\ 2007, Mer\'in et al.\ 2008).\ Cieza et al. (2007) found 
that all the known Classical T Tauri stars (CTTs), which are defined by having relatively 
strong nebular emission lines and thus being actively accreting, cluster around $\alpha_{\rm turnoff}$ 
$\sim$ $-1.0$ and $\lambda_{\rm turnoff}$ $\la$ $K_s$, whereas the Weak-line T Tauri stars 
(WTTs) exhibit a much larger spread of $\alpha_{\rm turnoff}$ and $\lambda_{\rm turnoff}$.\ 

\subsection{Fractional Dust Luminosity vs. $\lambda_{\rm turnoff}$}

The ratio of the circumstellar dust luminosity $L_{\rm dust}$ to stellar luminosity $L_{\star}$, 
which is also known as the fractional dust luminosity, was found to be correlated with the 
disk accretion activity (e.g.\ Kenyon \& Hartmann 1995; Muzerolle et al.\ 2003).\ In particular, for 
mildly flared dusty disks, $L_{\rm dust}$/$L_{\star}$ $\ga$ 0.1--0.2 cannot be simply explained by 
dust reprocessing of stellar radiation alone (Kenyon \& Hartmann 1995) 
but indicates that a significant amount of IR excesses may be contributed by self-radiation of an 
actively accreting disk, whereas YSOs with 0.001 $\la$ $L_{\rm dust}$/$L_{\star}$ $\la$ 0.1 are 
expected to be mostly evolved objects with weaker or no observable accretion activity 
(e.g.\ Cieza et al.\ 2007).\ Moreover, most gas-poor debris disks (systems which are dominated 
by second-generation dust produced by the collision of planetesimals) were found to have 
$L_{\rm dust}$/$L_{\star}$ well below 0.001 (e.g.\ Currie \& Kenyon 2009; Eiroa et al.\ 2013; 
Matthews et al.\ 2014; Su et al.\  2006; Trilling et al.\ 2008).\ 

\begin{figure}
   \centering
   \includegraphics[width=0.98\textwidth, height=0.35\textwidth]{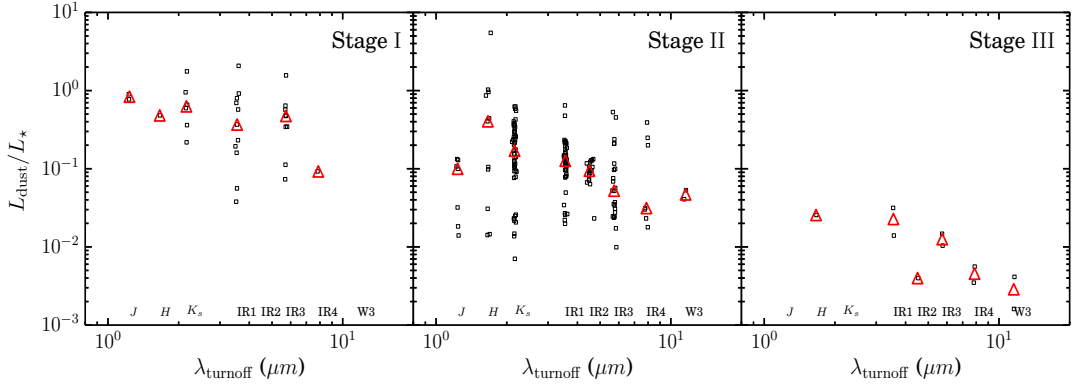}
   \vspace{-0.4cm}
   \caption{Distributions of the fractional dust luminosity $L_{\rm dust}/L_{\star}$ vs.\ $\lambda_{\rm turnoff}$.\
   YSOs of different evolutionary stages are plotted separately in different panels.\ The median 
   $L_{\rm dust}/L_{\star}$ at different $\lambda_{\rm turnoff}$ is represented as red triangles.
   }
   \label{fig10}
\end{figure}

We determined $L_{\rm dust}$ as an integral of the R06 model SED of the circumstellar 
dust (disk+envelope) that best fits the emergent IR excess emission, and $L_{\star}$ 
as $(R_{\star}/R_{\odot})^{2} (T_{\star}/T_{\odot})^{4}$, where $R_{\star}$ and $T_{\star}$ 
are the stellar radius and effective temperature.\ The distribution of YSOs of different Stages 
on the $L_{\rm dust}/L_{\star}$ vs. $\lambda_{\rm turnoff}$ diagram is 
shown in Figure \ref{fig10}.\ The Stages {\sc I} and {\sc III} YSOs are well separated at 
$L_{\rm dust}/L_{\star}$ $\sim$ 0.1, whereas the Stage {\sc II} YSOs have a range of 
$L_{\rm dust}/L_{\star}$ from $\sim$ 0.01 to 1.\ Moreover, there is a general trend that 
the median $L_{\rm dust}/L_{\star}$ decreases with increasing $\lambda_{\rm turnoff}$ for YSOs 
of different evolutionary stages, pointing to an inside-out disk clearing of at least the small 
dust grains.\

\subsection{Disk Inner Radius vs. $\alpha_{\rm excess}$}
\begin{figure}
   \centering
   \includegraphics[width=0.98\textwidth, height=0.8\textwidth]{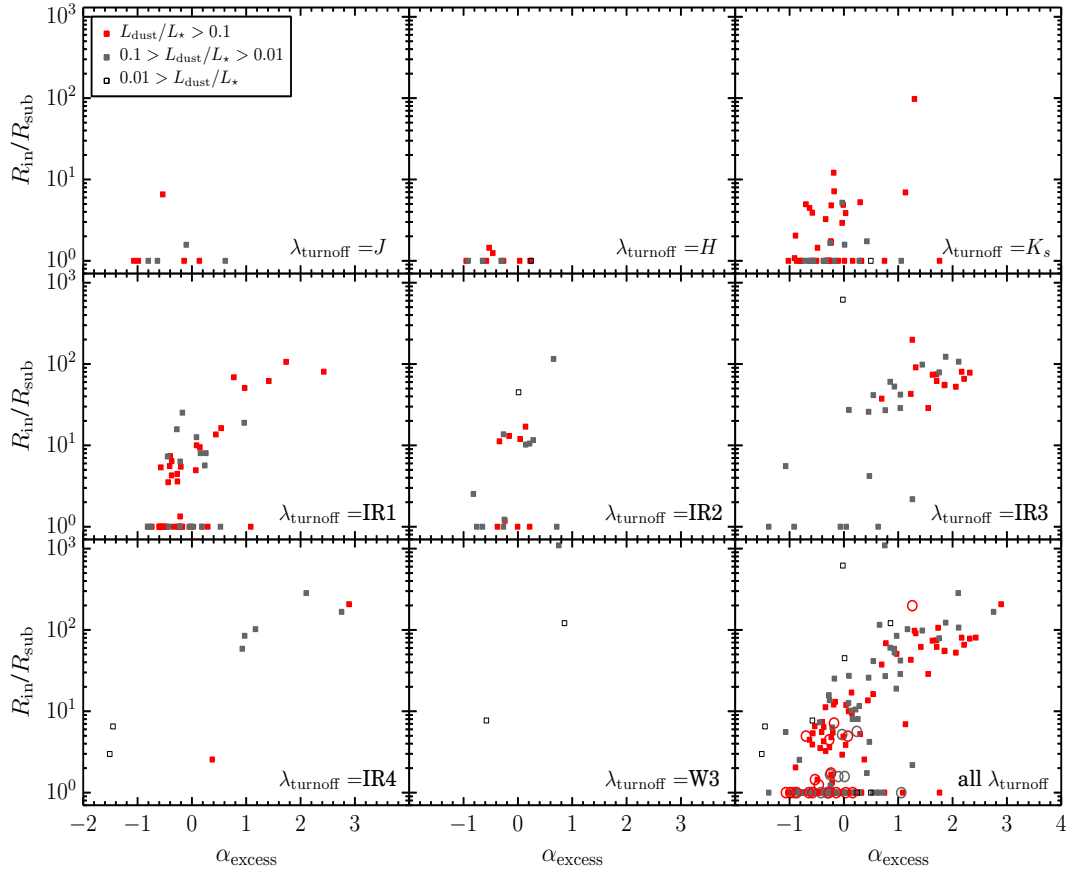}
   \vspace{-0.4cm}
   \caption{Distribution of disk inner radius $R_{\rm in}$ vs.\ $\alpha_{\rm excess}$ for different 
   $\lambda_{\rm turnoff}$.\ $R_{\rm in}$ is normalized to the dust sublimation radius $R_{\rm sub}$.\
   YSOs with different ranges of $L_{\rm dust}$/$L_{\star}$ are plotted with different symbols, 
   as indicated in the top left panel.\ Distribution for the full sample is shown in the bottom right panel, 
   where objects with 70$\mu$m detections are shown as red open circles.
   }
   \label{fig11}
\end{figure}

The inner radius of the dusty disk determines the highest temperature of dust grains orbiting around 
the central stellar source (e.g.\ Backman \& Paresce 1993), and thus can affect $\alpha_{\rm excess}$.\ 
Figure \ref{fig11} shows the relation between $\alpha_{\rm excess}$ and $R_{\rm in}/R_{\rm sub}$ with 
different $\lambda_{\rm turnoff}$, where $R_{\rm in}$ is the disk inner radius, and $R_{\rm sub}$ is the 
dust sublimation radius by assuming a sublimation temperature of 1600 K (R06).\ The bottom right 
panel of Figure \ref{fig11} shows the corresponding distribution for the full sample.\

There is a positive correlation between $R_{\rm in}/R_{\rm sub}$ and $\alpha_{\rm excess}$ at 
$R_{\rm in}/R_{\rm sub}$ $\ga$ 10 and $\alpha_{\rm excess}$ $\ga$ 0.0, irrespective of 
$L_{\rm dust}/L_{\star}$ and $\lambda_{\rm turnoff}$.\ A similar trend (not shown in the paper) 
also exists between $R_{\rm in}$ and $\alpha_{\rm excess}$ at $R_{\rm in}$ $>$ 0.5 AU 
and $\alpha_{\rm excess}$ $\ga$ 0.0.\ We note that a positive correlation was also 
found between disk inner radii (or hole radii) and disk masses for 35 c2d YSOs by 
Mer\'in et al.\ (2010).\

\subsection{Disk Flaring vs. $\alpha_{\rm excess}$}
\begin{figure}
   \centering
   \includegraphics[width=0.98\textwidth, height=0.8\textwidth]{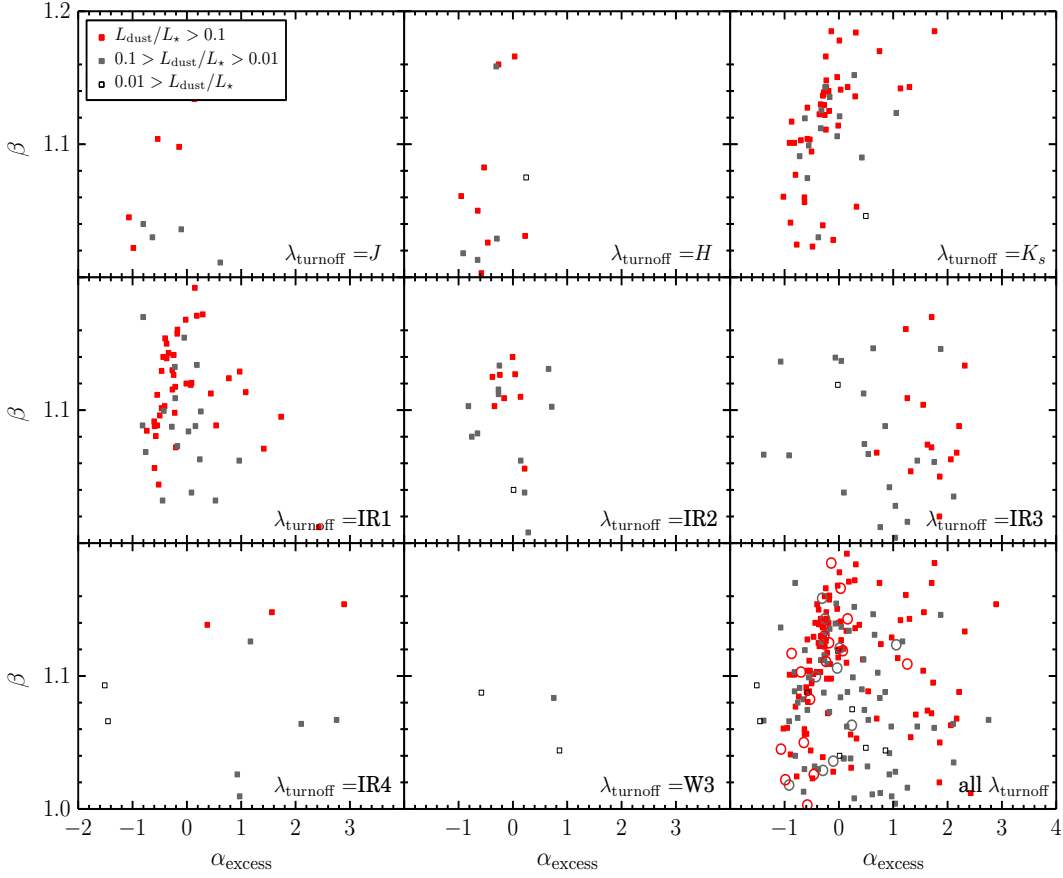}
   \vspace{-0.4cm}
   \caption{Distribution of disk flaring power $\beta$ vs.\ $\alpha_{\rm excess}$ for 
   different $\lambda_{\rm turnoff}$.\ As in Figure \ref{fig11}, YSOs with different ranges 
   of $L_{\rm dust}$/$L_{\star}$ are plotted with different symbols, as indicated in the top 
   left panel.\ Distribution of the full sample is shown in the bottom right panel, where 
   objects with 70$\mu$m detections are shown as red open circles.
   }
   \label{fig12}
\end{figure}

Compared to a completely flat disk geometry, a flaring geometry increases the disk area 
that confronts the stellar radiation at large radii, and thus enhances the mid- to far-IR emission 
(e.g.\ Kenyon \& Hartmann 1987; Chiang \& Goldreich 1997).\ The disk flaring power $\beta$ 
describes the radial gradient of the disk scale height $h$, i.e.\ $h(r)\propto r^{\beta}$, where $r$ 
is the cylindrical radius along the disk.\ Relationship between $\beta$ and 
$\alpha_{\rm excess}$ for our YSOs is shown in Figure \ref{fig12}.\ While no 
significant  correlation between $\beta$ and $\alpha_{\rm excess}$ was found for the overall sample, 
a majority of the disks with $\alpha_{\rm excess}$ $<$ 0.0 follow a trend that $\alpha_{\rm excess}$ 
increases with $\beta$, suggesting that the lack of a correlation between $R_{\rm in}/R_{\rm sub}$ 
and $\alpha_{\rm excess}$ for small disk inner radii can be in part attributed to the disk flaring.

\subsection{Discussion}

A variety of physical mechanisms have been invoked to explain the circumstellar disk evolution 
and clearing processes (e.g.\ Henning \& Meeus 2011; Williams \& Cieza 2011).\ The few 
commonly-considered mechanisms include viscous disk accretion (e.g.\ Hartmann et al.\ 1998; 
Lynden-Bell \& Pringle 1974; Shakura \& Sunyaev 1973), grain growth and dust settling 
(e.g.\ Dullemond \& Dominik 2005; Tanaka et al.\ 2005), photoevaporative dispersal 
(e.g.\ Alexander et al.\ 2006a,b; Gorti \& Hollenbach 2009; Hollenbach et al.\ 1994; Shu et al.\ 1993; ) 
and dynamical clearing by companion stars or planets (e.g.\ Artymowicz \& Lubow 1994; 
Kley \& Nelson 2012; Lubow \& D'Angelo 2006; Zhu et al.\ 2012).\ While all of these proposed 
processes may operate simultaneously, it is important to probe the dominant process(es) at 
different disk evolution stages.\

\subsubsection{From $\alpha_{\rm excess}$ to Disk Geometry}
The near- to mid-IR $\alpha_{\rm excess}$ is primarily affected by the inner disk clearing and 
outer disk flaring.\ In particular, the edge region of the optically thick inner disk, which is determined by either 
dust sublimation or some clearing processes, is illuminated directly by the stellar irradiation and 
thus contributes most of the hot dust emission excess, with the irradiation peak of this inner 
edge being shifted from near- to mid-IR as the disk is progressively cleared inside-out.\ 
In addition, as the disk evolves, dust settling or other clearing processes may result in a 
gradual reduction of disk flaring, which would in turn reduce the disk area that intercepts the 
stellar radiation and thus suppress the reprocessed cooler dust emission.\ 
Therefore, a progressively increasing disk inner edge is expected to increase 
$\alpha_{\rm excess}$, whereas a smaller flaring power in the outer disk can result in a 
smaller $\alpha_{\rm excess}$.\

Our results suggest that variation of $\alpha_{\rm excess}$ above $\sim$ 0.0 primarily 
reflects the variation of disk clearing radii, whereas variation of $\alpha_{\rm excess}$ 
below $\sim$ 0.0 is largely related to a variation of the disk flaring power.\ 
Disk flaring is only important in shaping the near-to mid-IR SEDs when $R_{\rm in}$ 
$\la$ 10$\times$$R_{\rm sub}$ ($>$ 0.5 AU for our sample).\
The lack of correlation between $\alpha_{\rm excess}$ and disk flaring power at $R_{\rm in}$ 
$\ga$ 10$\times$$R_{\rm sub}$ implies that either the outer disk geometry does not vary 
synchronously with the inside-out disk clearing processes or spectral slopes at $\lambda$ $\la$ 
24$\mu$m are not sensitive to the outer disk flaring.\ The small sample size of our disks (especially 
those with $\alpha_{\rm excess}$ $>$ 0.0) with detection at 70$\mu$m which is more sensitive 
to the outer disk flaring than shorter wavelengths (e.g.\ Sicilia-Aguilar et al.\ 2015) makes it hard to 
ascertain whether or not the outer disk flaring decreases or increases as the disk is cleared from the 
inside out.\ Recent studies of transitional disks in several nearby star-forming regions by 
Howard et al.\ (2013) and Keane et al.\ (2014) found that the continuum normalized 
[O {\sc i}] 63.18$\mu$m line luminosities, which traces the cool, outer disks, are suppressed 
by a factor of $\sim$ 2 on average with respect to the classical full disks, and this suppression was 
attributed to reduction of either the outer disk flaring or gas-to-dust ratio.

\begin{figure}
   \centering
   \includegraphics[width=0.98\textwidth, height=0.65\textwidth]{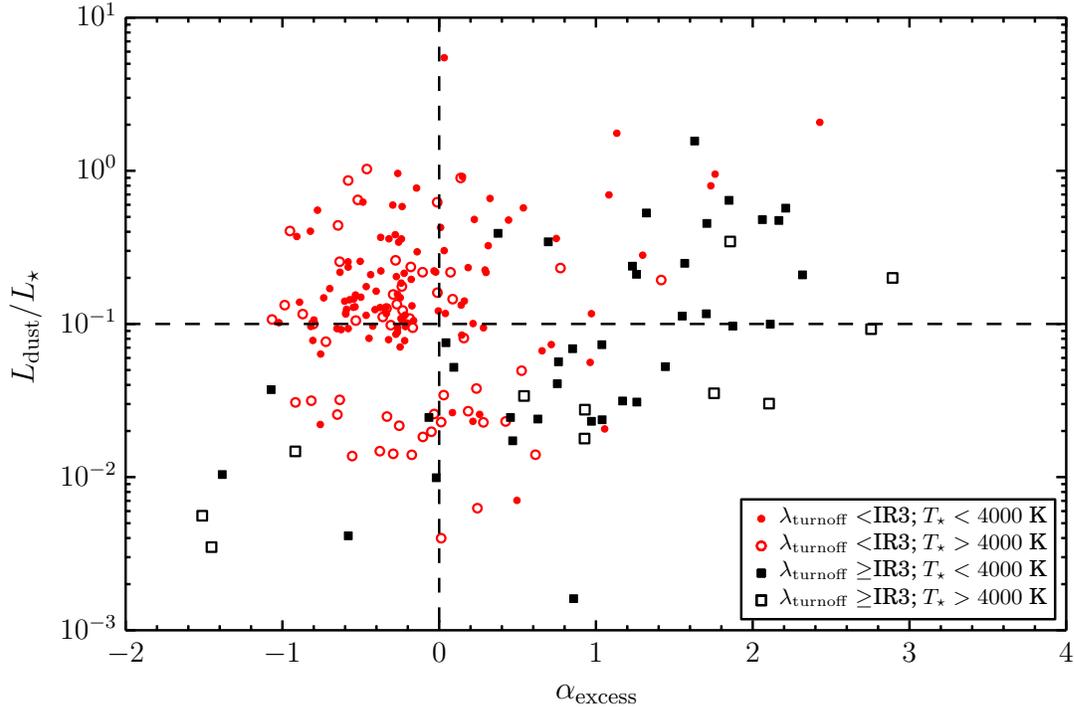}
   \vspace{-0.4cm}
   \caption{$\alpha_{\rm excess}$ is plotted against the fractional disk luminosities $L_{\rm dust}/L_{\star}$.\ 
   Disks with $\lambda_{\rm turnoff}$ $<$ and $\geq$ IR3 (transitional disks) are plotted as red open circles 
   and black filled  squares respectively.\ The horizontal dashed line separates the sample into disks with 
   $L_{\rm dust}/L_{\star}$ $>$ and $<$ 0.1, and the vertical dashed line separates the sample into disks 
   with $\alpha_{\rm excess}$ $>$ and $<$ 0.0.\ Most accreting disks were found to have $L_{\rm dust}/L_{\star}$ 
   $\geq$ 0.1.\ Transitional disks in the lower left part may be primarily cleared by photoevaporation, while 
   those in the upper right part may be dynamically cleared by giant planets.}
   \label{fig13}
\end{figure}

\subsubsection{Probing Disk Dispersal Processes with Transitional Disks}
There may be a variety of evolutionary paths from the optically thick full disks 
to optically thin to debris disks.\ Distinguishing different disk dispersal processes 
is crucial for understanding how the planetary systems formed from protoplanetary 
disks.\ The partially-cleared transitional disks, which have little or no excess emission 
in the near-IR ($\la 5\mu$m) and thus optically thin inner opacity holes but a significant 
excess at longer wavelengths (e.g.\ Brown et al.\ 2007; Calvet et al.\ 2005; Strom et al.\ 1989; 
Skrutskie et al.\ 1990), provide a unique opportunity to probe different disk clearing mechanisms 
because different mechanisms are expected to result in very different IR spectral slopes, 
disk luminosities, and accretion activities in the short transitional stages 
(e.g.\ Alexander et al.\ 2014; Cieza et al.\ 2010; Najita, Strom \& Muzerolle 2007).\ 

To open an inner opacity hole through photoevaporation, the disk viscous accretion rate 
has to fall below the photoevaporation rate (e.g.\ Alexander, Clarke \& Pringle 2006a; 
Owen et al.\ 2010), and once this happens, the full disks of gas and small dust grains can 
be quickly dissipated from the inside out in $\la$ 0.1 Myr which is an order of magnitude 
shorter than the typical disk lifetime.\ Besides a low fractional disk luminosity and steep IR 
spectral slope (e.g.\ $\alpha_{\rm excess} < 0.0$), another important consequence from  
photoevaporative clearing is that little or no accretion is expected once an inner hole is opened.\ 
In contrast, dynamical clearing by giant planets may sustain a small but still considerable 
amount of disk accretion across the inner opacity hole and a relatively high outer disk masses 
and luminosities and thus rising mid- to far-IR SEDs (e.g.\  Alexander 2008; Najita et al.\ 2007).\ 
In contrary to both photoevaporation and dynamical clearing, the pure grain growth and 
dust settling processes can result in an efficient depletion of small grains (and thus suppression 
of near- to mid-IR emission) from the inside out over time scales much smaller than 0.1 Myr 
(e.g.\ Dullemond \& Dominik 2005), with little direct influence on the accretion activity.\ 


All of our YSOs have $L_{\rm dust}/L_{\star}$ $>$ 10$^{-3}$, and 49 (23\%) have 
$\lambda_{\rm turnoff}\geq$ IR3 and thus can be classified as transitional disks.\ 
Recall that for our sample disks with $\lambda_{\rm turnoff}\geq$ IR3 exhibit a remarkably 
higher median and larger scatter of $\alpha_{\rm excess}$ than those with $\lambda_{\rm turnoff}<$ 
IR3 (Figure \ref{fig9}).\ The fraction of transitional disks in our sample is slightly higher yet 
still comparable to previous studies of nearby star clusters or star-forming regions 
(e.g.\ Currie et al.\ 2009; Dahm \& Carpenter 2009; Fang et al.\ 2009; Hern\'andez et al.\ 2007; 
Kim et al.\ 2009; Lada et al.\ 2006).\ The distribution of our sample on the $\alpha_{\rm excess}$ vs.\ 
$L_{\rm dust}/L_{\star}$ plane is shown in Figure \ref{fig13}, where the transitional 
disks are plotted as black squares ({\it filled} for those with $T_{\star} < 4000$ K, and 
{\it open} for those with  $T_{\star} > 4000$ K).\ Note that previous studies did not subtract 
the stellar photosphere emission for calculating the excess spectral index, which tends to  
underestimate the ``genuine'' $\alpha_{\rm excess}$.\ 

As is shown in Figure \ref{fig13}, the majority of the disks with $\lambda_{\rm turnoff} <$ IR3 
clustered toward the upper left corner, with $L_{\rm dust}/L_{\star}$ $\ga$ 10$^{-1}$ and 
$\alpha_{\rm excess} \la$ 0.0, whereas the disks with $\lambda_{\rm turnoff} \geq$ IR3 seem 
to follow a sequence from the upper right to the lower left, with none of them 
having $L_{\rm dust}/L_{\star}$ $>$ 10$^{-1}$ and $\alpha_{\rm excess} <$ 0.0.\
Most of the objects around the upper left corner are expected to have accreting full disks, and 
they are clearly separated from the population of transitional disks in Figure \ref{fig13}.\
A similar separation of transitional disks and full disks was also recently found 
by Sicilia-Aguilar et al.\ (2015) based on the relation between spectral indices and accretion rates.\
Among the objects with $\lambda_{\rm turnoff} <$ IR3, 14 (7\%) have $\alpha_{\rm excess} <$ 0.0 
and $L_{\rm dust}/L_{\star} \leq $ 0.003.\ These 7\% objects are consistent with being 
the so-called ``anemic'' (e.g.\ Lada et al.\ 2006) or ``homologously depleted'' 
(e.g.\ Currie \& Sicilia-Aguilar 2011) disks, which have detectable excess emission that 
decreases steadily at all wavelengths.\

Transitional disks toward the lower left corner of Figure \ref{fig13} may be more evolved 
than those toward the upper right.\ Among the 49 transitional disks, 41 (84\%) have 
$\alpha_{\rm excess} > 0.0$ and 8 (16\%) have $\alpha_{\rm excess} < 0.0$.\ Observations 
of UV continuum or recombination emission lines for all of our sample will be 
necessary for obtaining ongoing disk accretion rate.\ The accretion activities are 
known to be closely connected to the disk global properties, such as disk luminosities, 
masses and dust settling.\ If we instead use $L_{\rm dust}/L_{\star}$ to 
approximately discriminate disks with and without accretion activity at a dividing value 
$=$ 0.1, 17 (35\%) of the 49 transitional disks have $\alpha_{\rm excess} > 0.0$ and 
$L_{\rm dust}/L_{\star} > 0.1$, which may indicate the possibility of dynamical clearing by 
giant planets; Among the 32 (65\%) disks with $L_{\rm dust}/L_{\star} < 0.1$, 8 have 
$\alpha_{\rm excess} < 0.0$ and 24 have $\alpha_{\rm excess} > 0.0$, the low $L_{\rm dust}/L_{\star}$ 
probably indicates that these 32 disks are primarily cleared by photoevaporation.\ None of 
our transitional disks have $\alpha_{\rm excess} < 0.0$ and $L_{\rm dust}/L_{\star} > 0.1$, so grain growth and 
dust settling alone are probably not important hole-opening mechanisms (Ceiza et al.\ 2010).\ 
Furthermore, our finding that the median $\alpha_{\rm excess}$ of Stages {\sc I} 
and {\sc II} YSOs tend to increases with $\lambda_{\rm turnoff}$ also suggests that 
disk clearing is not primarily driven by grain growth which would otherwise 
results in a negative correlation between $\alpha_{\rm excess}$ and $\lambda_{\rm turnoff}$ 
(e.g.\ Dullemond \& Dominik 2005).\

\section{Summary}\label{sect: summ}
We have statistically explored the properties of the central stellar sources, 
the evolutionary stages, and the circumstellar disks for a sample of 211 Perseus YSOs by modeling  
the optical to mid-IR broadband SEDs with the R06 YSO evolution models.\
The median central stellar mass and age for the Perseus YSOs are $\sim$ 0.3 $M_{\sun}$ and 
$\sim$ 3.1 Myr respectively based on the Siess et al.\ (2000) PMS evolutionary models.\ 
About 81\% of our sample are classified as Stage {\sc II} objects which are characterized by  
having optically thick disks, $\sim$ 14\% are classified as Stage {\sc I} objects which are 
characterized by having significant infalling envelopes, and the remaining 5\% are classified 
as Stage {\sc III} objects with optically thin disks.\
Our primary results are summarized as follows.

\begin{itemize}

\item The evolutionary Stages as determined from the SED modeling have a general 
correspondence with the traditional spectral-indices-based Classes.\ In particular, 
$\sim$ 90\% of the Class {\sc II} YSOs fall into the Stage {\sc II} phase which is characterized 
by optically thick disks, and 75\% of the Class {\sc I} YSOs fall into the Stage {\sc I} phase which 
is characterized by significant infalling envelopes.\ Nevertheless, relating the Classes {\sc III}  and 
{\sc Flat} YSOs to specific evolutionary stages is uncertain.\ In particular, half of the Class {\sc III} 
YSOs fall into the Stage {\sc II} and the other half fall into the optically thin Stage {\sc III} phase, 
and half of the Class {\sc Flat} YSOs fall into the Stage {\sc I} and the other half fall into the 
Stage {\sc II} phase.\

\item We determined the turnoff wave band ($\lambda_{\rm turnoff}$) longward of which significant 
IR excesses with respect to the stellar photosphere level start to be observed and the excess spectral 
indices $\alpha_{\rm excess}$ at $\lambda$ $>$ $\lambda_{\rm turnoff}$.\ 
The median and standard deviation of $\alpha_{\rm excess}$ of the Stages {\sc I} and {\sc II} YSOs 
tend to increase with $\lambda_{\rm turnoff}$, especially at  $\lambda_{\rm turnoff} \geq$ IRAC 5.8$\mu$m.\
There is a general trend that the median fractional dust luminosity $L_{\rm dust}/L_{\star}$ 
decrease with increasing $\lambda_{\rm turnoff}$, pointing to an inside-out disk clearing of small dust grains.\
We found a positive correlation between $\alpha_{\rm excess}$ and disk inner radius $R_{\rm in}$, and 
a lack of correlation between $\alpha_{\rm excess}$ and disk flaring at $\alpha_{\rm excess} \ga 0.0$ 
and $R_{\rm in} \ga$ 10$\times$$R_{\rm sub}$, which indicates that, firstly, the near- to mid-IR spectral 
slopes primarily reflect the progressive disk clearing from the inside out once $R_{\rm in} \ga$ 
10$\times$$R_{\rm sub}$, secondly, the outer disk flaring either does not vary synchronously 
with the inner disk clearing processes or has little appreciable influence on the spectral slopes at 
wavelengths $\la$ 24$\mu$m. 

\item About 23\% (49) of our YSOs are classified as transitional disks, which have $\lambda_{\rm turnoff} 
\geq$ IRAC 5.8$\mu$m and $L_{\rm dust}/L_{\star}$ $>$ 10$^{-3}$.\
By using the $L_{\rm dust}/L_{\star}$ to approximately discriminate disks with and without accretion 
activity at a dividing value of 0.1, 35\% of the transitional disks have $\alpha_{\rm excess} > 0.0$ 
and $L_{\rm dust}/L_{\star} > 0.1$, implying the possibility of dynamical clearing by giant planets;
65\% have $L_{\rm dust}/L_{\star} < 0.1$, which is consistent with the expectation of 
photoevaporative clearing; None of our disks has $\alpha_{\rm excess} < 0.0$ and 
$L_{\rm dust}/L_{\star} > 0.1$, so grain growth and dust settling are probably not the 
driven mechanisms in disk clearing, in line with the trend that the median $\alpha_{\rm excess}$ 
increases, rather than decreases, with $\lambda_{\rm turnoff}$.

\end{itemize}

An indispensable diagnostic of the evolutionary stages of YSOs and their circumstellar disks 
is the current accretion rate, which is usually determined either from recombination lines 
or ultraviolet continuum excesses.\ Different disk clearing processes can lead to different disk 
accretion properties, the effect of which is especially prominent in transitional stages.\ 
Moreover, similar to many previous studies, our current work is heavily biased against 
the Stage {\sc III} YSOs with optically thin or anemic disks.\ To understand the disk evolution 
and dispersal processes, a systematic census of Stage {\sc III} YSOs and their disk accretion 
activity is imperative.\ Therefore, our future direction will include 1) a systematic spectroscopic 
followup of our YSOs with the Large Sky Area Multi-Object Fiber Spectroscopic Telescope 
(LAMOST; Cui et al.\ 2012) to place stringent constraints on the on-going accretion 
activity; 2) a wide-field time-series optical photometry across the whole Perseus region for 
an unbiased census of Stage {\sc III} disks with the PMO Xuyi 1.2-m Schmidt Telescope, in 
order to further probe the dominant disk dispersal mechanisms.\

\setlength{\tabcolsep}{5pt}
\footnotesize
\begin{longtable}{lllllllll}
\caption[]{Optical Photometry of Perseus YSOs}\label{tab1}\\
\hline\hline
ID$_{\rm c2d}$ & R.A.(J2000) & Decl.(J2000) & $g$ & $\sigma_{g}$ & $r$ & $\sigma_{r}$ & $i$ & $\sigma_{i}$\\
\hline
 & (deg) & (deg) & (mag) & (mag) & (mag) & (mag) & (mag) & (mag)\\
\endfirsthead
\hline\hline
ID$_{\rm c2d}$ & R.A.(J2000) & Decl.(J2000) & $g$ & $\sigma_{g}$ & $r$ & $\sigma_{r}$ & $i$ & $\sigma_{i}$\\
\hline
 & (deg) & (deg) & (mag) & (mag) & (mag) & (mag) & (mag) & (mag)\\
\hline
\endhead
\hline
\endfoot
\multicolumn{9}{l}{Notes: All the photometry was calibrated against the SDSS DR8 catalog.}\\
\endlastfoot
\hline
\multicolumn{9}{c}{$J$ band turnoff\footnote{The YSOs start exhibiting significant (3$\sigma$) IR excesses above the photosphere level longward of the turnoff wavebands.}}\\
\hline

c2dJ032852.2+304506  &   52.21736  &   30.75154  &     13.96  &      0.06  &      12.9  &      0.06  &            &          \\
c2dJ032854.6+311651  &   52.22763  &   31.28086  &            &            &     18.73  &      0.08  &      16.7  &      0.06\\
c2dJ032917.7+312245  &   52.32366  &   31.37917  &     15.55  &      0.06  &     13.82  &      0.05  &            &          \\
c2dJ033035.5+311559  &   52.64781  &   31.26627  &     18.78  &      0.07  &      18.0  &      0.07  &     17.05  &      0.06\\
c2dJ033330.4+311051  &   53.37669  &   31.18071  &     15.99  &      0.05  &     13.98  &      0.06  &            &          \\
c2dJ034157.4+314837  &   55.48934  &   31.81021  &     18.38  &      0.06  &     16.22  &      0.06  &      14.7  &      0.06\\
c2dJ034344.5+314309  &   55.93535  &   31.71926  &     18.69  &      0.06  &     17.19  &      0.06  &     14.52  &      0.06\\
\hline
\end{longtable}
\setlength{\tabcolsep}{3pt}
\footnotesize
\begin{longtable}{lllllllllllll}
\caption[]{SED Fitting Results of Perseus YSOs}\label{tab2}\\
\hline\hline
ID$_{\rm c2d}$ & ID$_{\rm model}$ & $\chi^{2}_{r}$ & $A_{V}$ & $D$ & $M_{\star}$ & Age$_{\star}$ & $L_{\star}$ & $L_{\rm dust}$ & $R_{\rm in}/R_{\rm sub}$ & Stage & $\alpha_{\rm turnoff}$ & $\alpha_{\rm excess}$\\
\hline
 & &  & (mag) & (kpc) & ($M_{\sun}$) & (10$^{5}$ yr) & (10$^{-2}$$L_{\sun}$) & (10$^{-2}$$L_{\sun}$) &  & & & \\
 (1) & (2) & (3) & (4) & (5) & (6) & (7) & (8) & (9) & (10) & (11) & (12) & (13)\\
\endfirsthead
\hline\hline
ID$_{\rm c2d}$ & ID$_{\rm model}$ & $\chi^{2}_{r}$ & $A_{V}$ & $D$ & $M_{\star}$ & Age$_{\star}$ & $L_{\star}$ & $L_{\rm dust}$ & $R_{\rm in}/R_{\rm sub}$ & Stage & $\alpha_{\rm turnoff}$ & $\alpha_{\rm excess}$\\
\hline
 & &  & (mag) & (kpc) & ($M_{\sun}$) & (10$^{5}$ yr) & (10$^{-2}$$L_{\sun}$) & (10$^{-2}$$L_{\sun}$) &  & & & \\
 (1) & (2) & (3) & (4) & (5) & (6) & (7) & (8) & (9) & (10) & (11) & (12) & (13)\\
\hline
\endhead
\hline
\endfoot
\multicolumn{13}{l}{Notes: (1): c2d ID;}\\
\multicolumn{13}{l}{(2): ID of the Best-fit model from Robitaille et al.\ (2006);}\\
\multicolumn{13}{l}{(3): Minimum reduced $\chi^{2}$;}\\
\multicolumn{13}{l}{(4): The most probable $V$-band interstellar extinction;}\\
\multicolumn{13}{l}{(5): The most probable heliocentric distance;}\\
\multicolumn{13}{l}{(6): The most probalbe stellar mass and the 95\% confidence interval;}\\
\multicolumn{13}{l}{(7): The most probable stellar age and the 95\% confidence interval;}\\
\multicolumn{13}{l}{(8): The best-fit stellar bolometric luminosity;}\\
\multicolumn{13}{l}{(9): Dust luminosity from integral of the model SED of disk+envelope that best fits the observed SED;}\\
\multicolumn{13}{l}{(10): The disk inner radius, in units of the dust sublimation radius, and the 95\% confidence interval;}\\
\multicolumn{13}{l}{(11): Evolutionary stage;}\\
\multicolumn{13}{l}{(12): Spectral indices at $\lambda >$ $\lambda_{\rm turnoff}$ for stellar photosphere-included IR SEDs.}\\
\multicolumn{13}{l}{(13): Spectral indices at $\lambda >$ $\lambda_{\rm turnoff}$ for stellar photosphere-subtracted IR SEDs.}\\
\endlastfoot
\hline
\multicolumn{13}{c}{$J$ band turnoff\footnote{The YSOs start exhibiting significant (3$\sigma$) IR excesses above the photosphere level longward of the turnoff wavebands.}}\\
\hline

c2dJ032852.2+304506  &  3000104  &   2.0  &  3.76  &    0.28  &  2.36$_{-0.22}^{+0.45}$  &  76.39$_{-23.22}^{+20.74}$  &         5439.09  &          173.77  &  1.7$_{-0.7}^{+0.7}$  &  ${\sc\rm II}$  &  -0.75  &  -0.63\\
c2dJ032854.6+311651  &  3013176  &   0.4  &  3.04  &    0.23  &  0.14$_{-0.04}^{+0.23}$  &  25.55$_{-21.9}^{+27.4}$  &            8.72  &            1.13  &  4.2$_{-3.2}^{+14.7}$  &  ${\sc\rm II}$  &  -0.57  &  -0.54\\
c2dJ032917.7+312245  &  3005510  &   0.5  &   5.7  &    0.26  &  2.22$_{-1.11}^{+0.93}$  &  76.42$_{-66.3}^{+22.48}$  &         2637.26  &          263.42  &  1.1$_{-0.1}^{+0.5}$  &  ${\sc\rm II}$  &  -0.95  &   -0.8\\
c2dJ033035.5+311559  &  3018343  &   6.7  &  0.79  &    0.24  &  1.02$_{-0.71}^{+1.83}$  &  8.92$_{-8.84}^{+31.75}$  &         2475.96  &         1910.17  &  83.9$_{-82.9}^{+237.7}$  &  ${\sc\rm I}$  &  -0.46  &  -0.14\\
c2dJ033330.4+311051  &  3003655  &   0.3  &   6.6  &    0.21  &  2.24$_{-0.26}^{+0.51}$  &  62.4$_{-26.81}^{+36.51}$  &         2561.06  &          339.83  &  1.0$_{-0.0}^{+-0.0}$  &  ${\sc\rm II}$  &   -0.8  &  -0.99\\
c2dJ034157.4+314837  &  3015603  &   1.3  &  7.69  &    0.28  &  2.08$_{-0.1}^{+0.13}$  &  93.61$_{-4.8}^{+5.3}$  &         1964.93  &          210.13  &  1.0$_{-0.0}^{+0.0}$  &  ${\sc\rm II}$  &  -0.86  &  -1.07\\
c2dJ034344.5+314309  &  3017083  &   2.5  &  3.27  &    0.28  &  3.11$_{-0.15}^{+0.3}$  &  16.84$_{-3.08}^{+59.29}$  &         6740.19  &          123.32  &  1.5$_{-0.5}^{+0.7}$  &  ${\sc\rm II}$  &  -0.13  &   -0.1\\
\hline
\end{longtable}

\begin{acknowledgements}

We thank the anonymous referee for his/her helpful comments that improved this manuscript.\
We acknowledge the support of the National Natural Science Foundation of China grant \#11390373.
HXZ acknowledges support from China Postdoctoral Science Foundation under Grant No. 2013M530008, 
and CAS-CONICYT Postdoctoral Fellowship, administered by the Chinese Academy of Sciences 
South America Center for Astronomy (CASSACA).\ MF acknowledges the NSFC under grant 11203081.\

\end{acknowledgements}

\label{lastpage}

\end{document}